\documentclass[]{article}
\usepackage[colorlinks]{hyperref}
\usepackage{framed}
\usepackage{graphicx}
\usepackage{subfigure}
\usepackage{appendix}
\usepackage{amssymb,amsfonts}
\usepackage{amsmath}
\usepackage{caption}
\usepackage{bbold}

\newcommand{\tr}{\mbox{$\mathrm{Tr}$}}

\title{Calogero-Moser models with internal degrees of freedom revisited}
\author{Katarzyna Kowalczyk-Murynka\footnote{email: kkowalczyk@cft.edu.pl} \hskip 5pt and Marek Ku\'s\footnote{email: marek.kus@cft.edu.pl} \\
\textit{Center for Theoretical Physics, Polish Academy of Sciences} \\
\textit{Al. Lotnik\'ow 32/46, 02-668 Warszawa, Poland}}

\begin{document}
\renewcommand{\figurename}{Fig.}
\renewcommand{\tablename}{Tab.}

\maketitle

\begin{abstract}
\noindent
	We discuss various examples of classical Calogero-Moser models with internal degrees of freedom. These models besides of having some attractive properties, like the complete integrability, are of interest eg., in studying spectral properties of quantum chaotic systems. The role of internal degrees of freedom is important in at least two aspects. Firstly, they come in play as dynamically evolving couplings  between repelling pairs of eigenvalues, hence they influence the speed of eigenvalue dynamics. Secondly their initial values determine the  "reachable sets" of couplings between particular pairs accessible during the evolution. The considered models are studied in a framework of matrix dynamics in a unified way based on a reduction of a linear model in an extended phase-space. Such an approach enables showing an equivalencies among various types of similar models employing "vectorial degrees of freedom" and constructing new systems of similar type.   
\end{abstract}
\section{Introduction}
In a series of papers published nearly half a century ago, Calogero presented explicit solutions of a quantum one-dimensional many-body system with a Hamiltonian
\cite{calogero69,calogero69a,calogero71}
\begin{equation}\label{calogero}
\mathcal{H}_{CM}=\frac{1}{2}\sum_{i} p_i^2 + \frac{g^2}{2}\sum_{i\neq j}\frac{1}{(x_i-x_j)^2}. 
\end{equation}
The system happens to be (quantum) integrable \cite{olshanetsky83}, and its classical counterpart, for which (\ref{calogero}) is the Hamilton function, is integrable classically \cite{olshanetsky81}. The Calogero-Moser (CM) system (\ref{calogero}) is a member of a very limited family of fully integrable systems of similar type. For them the two body interaction $V(x)=x^{-2}$ in (\ref{calogero}) can be exchanged for $V(x)=x^{-2}+\omega^2x^2$ \cite{calogero71}, $V(x)=a^2\sin^{-2}(ax)$ \cite{sutherland71,sutherland72}, $V(x)=a^2\sinh^{-2}(ax)$ \cite{calogero75},  and $ V(x)=a^2\mathcal{P}(aq) $, where $ \mathcal{P} $ is the Weierstrass function \cite{olshanetsky81}.     

Another way of generalizing (\ref{calogero}) is to introduce internal degrees of freedom of the interacting particles \cite{gibbons84,ha92,hikami93,minahan93,krichever95,kawakami94}. 
\begin{equation}
\mathcal{H}=\frac{1}{2}\sum_{i} p_i^2 + \frac{1}{2}\sum_{i\neq j}\frac{g_{ij}^2}{(x_i-x_j)^2} \label{HCMgij0},
\end{equation} 
where the potential part of the Hamilton function depends \textit{via} variable couplins $g_{ij}$ on dynamical variables characterizing additional degrees of freedom. In what follows we remind how to derive such models by an appropriate reduction procedure from a simple free or harmonic oscillator motion in a larger space. Such an approach puts all the known models on the same ground and enables identification of equivalent systems, as well as introduction of new similar ones.

In the study of spectral properties of quantum mechanical chaotic systems one employs generalized Calogero-Moser (or, Sutherland-Moser) systems to investigate universal spectral properties. They are exhibited as features of equilibrium dynamics of eigenvalues (where the fictitious time variable is proportional to a parameter governing the chaos in the system in question) \cite{haake19}. Internal degrees of freedom control the repulsion between eigenvalues, but since they themselves undergo a dynamical evolution, the strength of repulsion, and consequently, time scales of evolution, depend on details of a system (eg.\ its symmetries).  

From the same point of view, ie., the parametric dynamics of eigenvalues in quantum chaotic systems, it is interesting to identify "reachable sets" of values of the internal degrees of freedom for different models and various initial conditions - a classical problem of the control theory. For the level dynamics it enables to decide which couplings play a crucial role in the level repulsion.

The paper is organized as follows. In Section~\ref{matrix_dyn} we review the reduction approach to derivation of various types of Calogero-Moser systems. Sections~\ref{sec:evolution} and~\ref{sec:reach} we devote to the dynamics and reachable sets of internal degrees of freedom. In Section~\ref{sec:vectorial} we discuss CM-models in terms of vectorial degrees of freedom and in~\ref{sec:beyondcm} we show how to construct a new model of this type. In Section~\ref{sec:other} we show equivalencies among several models known in the literature and conclude with some outlook in Section~\ref{sec:conclusions}. Some details of calculations are presented in the Appendix. 

\section{The generalized Calogero-Moser system as a reduction of linear matrix dynamics}\label{matrix_dyn}
The original Calogero-Moser system  as well as all generalized ones with integral degrees of freedom can be obtained as a reduction of a linear Hamiltonian system on the space of matrices \cite{moser75} (in this paper we consider only Hermitian and real-symmetric matrices without additional symmetries, but further generalizations are conceivable \cite{kus97,huckleberry08}) To this end recall that a standard setting for Hamiltonian dynamics is that of a phase space, i.e., a differentiable manifold equipped with a symplectic form (a closed and non-degenerate two form) and a Hamiltonian function determining dynamics \cite{abraham78}. In most common case the phase space is the cotangent bundle $T^*\mathcal{M}$ of a configuration space $\mathcal{M}$. The construction simplifies further if $\mathcal{M}$ is a linear space. Then we have $T^*\mathcal{M}=\mathcal{M}\times\mathcal{M}$.    

In what follows we take as a configuration space a linear space of  $N\times N$ complex Hermitian matrices: $\mathcal{M}=\{X\in M_N(\mathbb{C}): X^\dagger=X \}$. According to what is written above, the corresponding phase space is thus  $\mathcal{M}\times\mathcal{M}=\{(X,Y): X^\dagger=X,Y^\dagger=Y\}$, and the canonical symplectic form reads  
\begin{equation}\label{symplecticform}
	\omega=\sum_{i,j}dY_{ij}\wedge dX_{ji}=\tr(dY\wedge dX),
\end{equation}
	
The Poisson brackets are constructed in a standard way. For a smooth function $f$ on the phase-space we define the corresponding vector field $\mathbf{X}_f$ \textit{via}
\begin{equation}\label{hamfield}
\omega(\mathbf{X}_f,\mathbf{Z})=-dF(\mathbf{Z}),
\end{equation}
for na arbitrary vector field $\mathbf{Z}$. A straightforward calculation shows that
\begin{equation}\label{hamfield1}
\mathbf{X}_f=\tr\left(\frac{\partial f}{\partial Y}\frac{\partial}{\partial X}- \frac{\partial f}{\partial X}\frac{\partial}{\partial Y}\right),	
\end{equation}	 
where the differentiation over matrix is meant as entry-wise.

The Poisson bracket of two functions $f$ and $g$ is now defined as
\begin{equation}\label{poissonbracket}
\left\{f,g\right\}=\omega(\mathbf{X}_f,\mathbf{X}_g),
\end{equation}
which, upon (\ref{symplecticform}) and (\ref{hamfield1}), gives	
\begin{equation}\label{PBmatr}
\left\{f,g\right\}=\frac{\partial f}{\partial X}\frac{\partial g}{\partial Y}-
\frac{\partial g}{\partial X}\frac{\partial f}{\partial Y}.	
\end{equation}
	
The equation of motion for an arbitrary phase-space function $f$ generated by a Hamiltonian function $\mathcal{H}$ reads, in terms of the Poisson bracket, as
\begin{equation}\label{hameqs}
\dot{f}=\left\{f,\mathcal{H}\right\}.
\end{equation}

\subsection{Linerar Hermitian matrix dynamics}\label{linred}	
Let us look at the dynamics given by the following Hamiltonian:
\begin{equation}\label{HCM_mat}
\mathcal{H}=\frac{1}{2}\tr YY^\dagger=\frac{1}{2}\tr Y^2.
\end{equation}
The equations of motion 
\begin{equation}
\dot{X}=Y, \quad  \dot{Y}=0,
\end{equation}
have an obvious solution,
\begin{equation} \label{lin_mat_dyn}
X(t)=X_0+t\cdot Y_0, \quad Y(t)=Y_0.
\end{equation}
Let us now choose new, time-dependent coordinates in which $X(t)$ remains diagonal (we assume that we start from a diagonal $X(0)=:X_0$), i.e. we diagonalize $X(t)$ at each instant of time. This leads to:
\begin{equation} \label{diagonal}
U(t)\left(\begin{array}{c} X(t) \\ Y(t) \end{array}\right)U(t)^\dagger =\left(\begin{array}{c} D(t) \\ V(t) \end{array}\right),
\end{equation}
 The equations of motion in this parametrization read:
\begin{eqnarray}
\dot{D}&=& V+[A,D]   \label{Dmdot} \\
	\dot{V}&=& [A,D] \label{Vmdot} \\
	\dot{L}&=& [A,L] \label{Lmdot} \\
	\dot{U}&=& AU    \label{Udot} 
	\end{eqnarray}
where $D(t)$ is diagonal, $A=\dot{U}U^{\dagger}$ and $L:=[D,V]$ is anti-Hermitian with $0$ on the diagonal. For the matrix entries we get:
	\begin{eqnarray}
	\dot{D}_{ii}&=& V_{ii} \label{Dii}\\
	0&=& V_{ij}+[A,D]_{ij} \label{D_offdiag}\\
	\dot{V}_{ii}&=& [A,D]_{ii} \label{Vdiag}\\
	\dot{L}_{ij}&=& [A,L]_{ij} \label{Lij}
	\end{eqnarray}
The entries of $A$ can be expressed \textit{via} $V$ and $D$ entries (\ref{D_offdiag}), and for $i\neq j$ we have $L_{ij}=(D_{ii}-D_{jj})V_{ij}$. 

$L_{ii}=0$ is not automatically true in general, but for the initial conditions ($X_0$ - diagonal) and for the chosen parametrisation (eigenvalues of X(t)) it just is and we do not need to set this value. We may, on the other hand, set $A_{ii} = 0$ - this is true for a specific gauge.\footnote{For a discussion of this condition, especially in connection with a canonical formulation of the problem, see \cite{haake19}, Ch.~11.}

 Finally we get a closed set of equations for $(D_{ii},V_{ii},L_{ij})$. To make a direct connection to the original Calogero-Moser system (\ref{calogero}) let us rename $x_i:= D_{ii}$, $p_i := V_{ii}$. The resulting system of equations reads
	\begin{eqnarray}
	\dot{x}_{i}&=& p_{i} \label{xdot}\\
	\dot{p}_{i}&=& \sum_{k\neq i} \frac{-2L_{ik}L_{ki}}{(x_{i}-x_{k})^3} \label{pdot}\\ 
	\dot{L}_{ij}&=& \sum_{k \neq i,j} L_{ik}L_{kj}\left(\frac{1}{(x_{i}-x_{k})^2}-\frac{1}{(x_{j}-x_{k})^2}\right). \label{Ldot}
\end{eqnarray}
Equations (\ref{xdot})-(\ref{Ldot}) are still Hamilton equations of motion derived from the old Hamilton function,  $\mathcal{H}=\frac{1}{2}\tr Y^2=\frac{1}{2} \tr V^2$ expressed in the new parametrization,

\begin{equation}\label{CML}
	\mathcal{H}_{CM,L}=\frac{1}{2}\sum_{i} p_i^2 - \frac{1}{2}\sum_{i\neq j}\frac{L_{ij}L_{ji}}{(x_i-x_j)^2}, 
\end{equation} 
describe the dynamics of $N$ interacting particles on a line. The Poisson brackets derived from (\ref{PBmatr}),
\begin{equation}\label{PBLP}
\left\{f,g\right\} =\frac{\partial f}{\partial {\bf x}}\cdot \frac{\partial g}{\partial {\bf p}}-\frac{\partial g}{\partial {\bf x}}\cdot \frac{\partial f}{\partial {\bf p}}+\tr\left( L\left[ \left( \frac{\partial f}{\partial L}\right)^T,\left( \frac{\partial g}{\partial L}\right) ^T\right]\right),
\end{equation}
where $^T$ denotes the matrix transpose \cite{huckleberry01}. Thus the Poisson bracket is a sum of two parts, a "canonical" one involving $ \mathbf{x} $ and $ \mathbf{p} $ and a Poisson bracket for anti-Hermitian matrices $L$ forming the $\mathfrak{su}(N)$  Lie algebra\footnote{In fact the Poisson brackets for $L$ are determined from the so called Kirillov-Kostant-Souriau symplectic structure on coadjoint orbits of a Lie group (in this case $SU(N)$). \cite{kirillov04}.}

Explicitly,
\begin{equation}\label{pb1}
\left\{ x_m,p_n\right\} =\delta _{mn},\quad 
\left\{ L_{mn},L_{ij}\right\} =\delta_{in}L_{mj}-\delta_{mj}L_{in}, 
\end{equation}
and all other Poisson brackets vanish.
The interaction in (\ref{CML}) is always repulsive, because the $L$ matrix is anti-Hermitian and therefore,
\begin{equation}
-L_{ij}L_{ji}=|L_{ij}|^2. \nonumber
	\end{equation}

Let us conclude this part by few short subsections concerning properties of the obtained dynamical system and its analogues.

\subsubsection{Integrals of motion}\label{sec:intmot}

From (\ref{Vmdot}) and (\ref{Lmdot}) due to the commutator form of both equations it is easy to infer integrals of motion of the system
\begin{equation}\label{intmotion}
I_{k_1,k_2...,k_M}=\tr({L^{k_1}V^{k_2}\cdots L^{k_{M-1}}V^{k_M}}),
\end{equation}
where $k_1,...k_M$ are arbitrary natural numbers and $V$ is expressed in terms of $x_i, p_i$, and $L_{ij}$. For the set of independent integrals and their number consider \cite{mnich93}. 

\subsubsection{External harmonic potential}

The same procedure applied to $H(X,Y)=\frac{1}{2}\tr(X^2+Y^2)$ leads to a silimar system of repelling particles, but in an external harmonic potential, 
\begin{equation}
\mathcal{H}_{C,L}=\frac{1}{2}\sum_{i} p_i^2+x_i^2 - \frac{1}{2}\sum_{i\neq j}\frac{L_{ij}L_{ji}}{(x_i-x_j)^2} \label{CL}
\end{equation}

\subsubsection{Sutherland model}

If, instead of (\ref{HCM_mat}), we chose as a Hamilton function $H=\tr(XY)^2$, and as phase space $\{(X,Y): X^\dagger=X^{-1},(XY)^\dagger=-XY\}$, the solution reads $X(t)=X_0\exp(X_0Y_0\cdot t)$, $Y(t)=\exp(-X_0Y_0\cdot t)$. In this case $X$ is unitary, and we denote its eigenphases by $x_k$ (i.e., the phases of its eigenvalues 
$\lambda_k=\exp(ix_k)$). Repeating the same procedure of reduction as before we arrive at a Hamilton function of the same form analogous to 
(\ref{CML}) , but with the interaction potential $\sin^{-2}x$ instead of the inverse quadratic one (the Sutherland model \cite{sutherland71,sutherland72}). 
%

\subsubsection{Unitary and orthogonal settings}\label{un-orth}

It is obvious that if the initial conditions $(X_0,Y_0)$ are real and symmetric, the dynamics stays in the subspace of real symmetric matrices, and such an evolution can be treated as a special case of (\ref{xdot})-(\ref{Ldot}). On the other hand we can construct it independently if we choose as a phase space $T^*\mathcal{M}=\mathcal{M}\times\mathcal{M}=\{(X,Y): X^T=X,Y^T=Y\}$, i.e. the cotangent bundle to the space of real symmetric matrices.

The reduction procedure in this case is the same as presented in (\ref{linred}). The resulting equations are exactly (\ref{xdot})-(\ref{Ldot}), and the Hamilton function is the same as (\ref{CML}) (or (\ref{CL}) in the harmonic case). The only differences are that $X(t),Y(t)$ are real, symmetric matrices, $L(t)$ is real, antisymmetric and, instead of a unitary matrix $U(t)$, we use an orthogonal diagonalizing matrix $O(t)$, so $A(t)=\dot{O}(t)O^T(t)$. 

The Poisson brackets have the form (\ref{PBLP}) but with the part coming from   Kirillov-Kostant-Souriau symplectic structure applied to the $SO(N)$ instead of the $U(N)$ group \cite{kus97,haake19}, reflecting the fact that now $L$ belongs to the $\mathfrak{so}(N)$ algebra. Explicitly,
\begin{equation}\label{pbso}
\{L_{ij},L_{kl}\}=\frac{1}{2}\left(\delta_{kj}L_{il}-\delta_{il}L_{kj}-\delta_{jl}L_{ik}+\delta_{ik}L_{lj}\right)
\end{equation}

We shall call this case the orthogonal setting and discuss it below in Section~\ref{ort}. 

In investigation of spectral properties of quantum chaotic systems different settings (symmetry classes) correspond to time-reversal invariant (orthogonal setting) and noninvariant systems \cite{haake19}.     

\subsubsection{Gauge invariance}\label{ginv}
The matrix dynamics described by (\ref{diagonal}) and consequently by (\ref{xdot})-(\ref{Ldot}) admits a kind of gauge invariance resulting from the fact that a diagonalizing matrix $U(t)$ is not unique. If $U(t)$ does the job, so does $EU(t)$, where
\begin{equation}
 E=diag(e^{i\phi_1},e^{i\phi_2},...,e^{i\phi_N}). \label{gauge}
\end{equation}
In the orthogonal setting $\phi\in\left\{0,\pi\right\}$, which means that $E\in\mathbb{Z}_2^{\times N}$, while in the unitary setting $\phi\in[0,2\pi)$ and $E\in U(1)^{\times N}$. 
This means that we can treat the evolution of $ L $ that, according to (\ref{Lmdot}), (\ref{Udot}) reads $ L(t)=U(t)L(0)U^\dagger(t) $,
as pertaining to an equivalence class $[L] = \left\{L^{\prime}=ELE^\dagger\right\}$, where $E$ belongs to an appropriate set.
$L$ and $L^{\prime}$ belong to the same equivalence class, i.e, $[L]=[L^{\prime}]$, if and only if
\begin{equation}
L^{\prime}_{ij}=L_{ij}e^{i(\phi_i-\phi_j)}.
\end{equation}
Writing $L_{ij}=|L_{ij}|e^{i\varphi_{ij}}$ and $L^{\prime}_{ij}=|L^{\prime}_{ij}|e^{i\varphi^{\prime}_{ij}}$ $i<j$ we clearly get $ |L_{ij}|=|L^{\prime}_{ij}| $, but there must also  exist such $\phi_1,\phi_2,...,\phi_N\in[0,2\pi)]$, that for all $i<j$:
\begin{equation}
\varphi^{\prime}_{ij}=\varphi_{ij}+\phi_i-\phi_j. \label{gauge1}
\end{equation}
If such N phases exist, every triple $(i,j,k)$ of indices must satisfy:
\begin{equation}
\varphi^{\prime}_{ij}+\varphi^{\prime}_{jk}+\varphi^{\prime}_{ki}=\varphi_{ij}+\varphi_{jk}+\varphi_{ki} 
=: \Phi_{ijk}.
\label{gauge2}
\end{equation}
On the other hand if (\ref{gauge2}) is satisfied, $\phi_1,\phi_2,...,\phi_N$ are well defined by (\ref{gauge1}). Indeed, let us define $ \alpha_i = \varphi_{i,i+1}^{\prime}-\varphi_{i,i+1}$, $i=1,2,\dots,N-1 $. Then, since $L$ and $L^\prime$ are anti-Hermitian, i.e., $\varphi_{ij}=-\varphi_{ji}$, $\varphi_{ij}^\prime=-\varphi_{ji}^\prime$, we get from (\ref{gauge2}) $\varphi_{j,k}^{\prime}-\varphi_{j,k}=\alpha_j+\ldots+\alpha_{k-1}$ for $ j<k-1 $. Now defining $ \phi_k = \alpha_1+\alpha_2+\ldots+\alpha_{k-1} $ we easily obtain (\ref{gauge1}). Therefore (\ref{gauge2}) is both sufficient and necessary for two matrices to be gauge equivalent.

\subsubsection{Time-reversal invariance}
There is another symmetry of the considered system, $ L^{\prime}=-L $,  not connected with (\ref{gauge}), since it corresponds to anti-unitary rather than unitary diagonal transformation, for which  
$\varphi^{\prime}_{ij}+\varphi^{\prime}_{jk}+\varphi^{\prime}_{ki}=-(\varphi_{ij}+\varphi_{jk}+\varphi_{ki})$. This is precisely the time reversal transformation. Yet the Hamilton function is time-reversal invariant, so the change $L\rightarrow -L$ corresponds to the same dynamics. 

\subsubsection{$N=2$ case}
In the case of $N=2$ the right-hand side of (\ref{Ldot}) vanishes, hence $L_{12}$ (the only relevant element of $L$) is constant and the presented generalization is trivial, i.e., it is impossible to go beyond the ordinary system (\ref{calogero}). This simple observation leads to a conclusion that the three-body generalized Calogero-Moser system is the smallest one for which the $L$ variables play a role.

\subsubsection{The ordinary Calogero-Moser system as a special case of the generalized one}\label{init}
We have just seen that for $N=2$ the generalized CM system reduces to the ordinary one. In fact for an arbitrary $N$ the ordinary CM system is a special case obtained from the generalized model by imposing appropriate initial conditions. To show it, let us take as the initial condition an anti-Hermitian, off-diagonal matrix

%
%
\begin{equation}
L(0)=-ig(\mathbb{1}-|e\rangle\langle e|), \label{L0}
\end{equation}
where $\langle e|=(1,1,....,1)$ in the eigenbasis of $X(0)$. 
Since $L(0)=[X(0),Y(0)]$ we get
\begin{equation}
L(t)=[D(t),V(t)]=[U(t)X(t)U^\dagger (t),U(t)Y(t)U^\dagger (t)]=U(t)[X(0),Y(0)]U^\dagger (t)=U(t)L(0) U^\dagger (t).
\end{equation}
For the initial condition (\ref{L0}):
\begin{equation}
L(t)=-ig U(t)(\mathbb{1}-|e\rangle\langle e|)U^\dagger (t)=-ig(\mathbb{1}-|f\rangle\langle f|),
\end{equation}	
where $|f\rangle= U(t)|e\rangle$, $\langle f|=(f^*_1,f^*_2,...,f^*_N)$. 
On the other hand, $L(t)=[D(t),V(t)]$ with a diagonal $D(t)$, so for all $t$ its diagonal elements vanish, 
\[
L_{ii}(t)=-ig(1-|f_i|^2)=0,
\]
which means  $\langle f|=(e^{i\phi_1},...,e^{i\phi_N})$. Thus, the off-diagonal elements read $L_{ij}(t)=ig\exp(i(\phi_i-\phi_j))$, but the phase factors $\exp(i\phi_i)$ can be eliminated by a choice of gauge. Indeed, $|f\rangle=E|e\rangle$, where $E = diag((e^{-i\phi_1},...,e^{-i\phi_N})$. On the other hand $|f\rangle=U|e\rangle$, so $E|e\rangle=U|e\rangle$. But if $U$ diagonalizes $X$, so does $E^\dagger U$ which preserves $|e\rangle$. We therefore can choose such a gauge in which $|e\rangle$ is preserved and consequently $L(t)=L(0)$, with $|L_{ij}|^2=g^2$ and the dynamics on this manifold is governed by the Hamilton function (\ref{calogero}).

One can ask whether this solution is unique, or are there any other, non-trivially different $L_0$ matrices recovering the same ordinary CM dynamics. In particular one can search for initial conditions corresponding to the Calogero-Moser system with different coupling constants:
\begin{equation}
\mathcal{H}_{CM,g}=\frac{1}{2}\sum_{i} p_i^2 + \frac{1}{2}\sum_{i\neq j}\frac{g_{ij}^2}{(x_i-x_j)^2} \label{HCMgij}
\end{equation} 
 The question of general conditions for $|L(t)|=|L_0|$ in the $(X,Y)$ phase space or, in other words, to the embedding of the ordinary Calogero-Moser system as a submanifold in the generalized space will be addressed in Section (\ref{rorbits})).
 
\section{Evolution of the internal degrees of freedom}\label{sec:evolution}
 
\subsection{The influence of the internal degrees of freedom on the $\mathbf{(x,p)}$ dynamics}
The equations of motion of the ordinary N-body Calogero-Moser system with $g_{ij}$ coupling constants (\ref{HCMgij}) are the following:
\begin{eqnarray}
\dot{x}_{i,g} &=& p_{i,g}, \\
\dot{p}_{i,g} &=& 2\sum_{k\neq i} \frac{g_{ik}^2}{(x_{i,g}-x_{k,g})^3},
\end{eqnarray}
while the equations of motion for $(x_i,p_i,L_{ij})$ variables which stem from (\ref{CML}):
	\begin{equation}
	\dot{x}_{i,L}=p_{i,L},\hspace{5mm}
	\dot{p}_{i,L}=\sum_{k\neq i} \frac{-2L_{ik}L_{ki}}{(x_{i,L}-x_{k,L})^3},\hspace{5mm} 
	\dot{L}_{ij}=\sum_{k \neq i,j} L_{ik}L_{kj}\left(\frac{1}{(x_{i,L}-x_{k,L})^2}-\frac{1}{(x_{j,L}-x_{k,L})^2}\right).
	\end{equation}
Assuming the two system are prepared in analogous states: 
\begin{equation}
x_{i,g}(0)=x_{i,L}(0),\hspace{5mm}
p_{i,g}(0)=p_{i,L}(0),\hspace{5mm}
g_{ij}=|L_{ij}|(0),
\end{equation}
we expand $(x_{i,g}-x_{i,L})(t)$ in a Taylor series around $t=0$ to see how quickly the $L_{ij}$ variables become significant for the motion in physical space:
\begin{eqnarray} \label{xshort}
(x_{i,g}-x_{i,L})(t)&=&(x_{i,g}-x_{i,L})(0)+t(\dot{x}_{i,g}-\dot{x}_{i,L})(0)+\frac{t^2}{2}(\ddot{x}_{i,g}-\ddot{x}_{i,L})(0)+\frac{t^3}{6}(x^{(3)}_{i,g}-x^{(3)}_{i,L})(0)+\mathcal{O}(t^4)= \nonumber \\
 &=& -\frac{t^3}{6}\left(\sum_{k\neq i} \frac{6g_{ik}^2\dot{x}_{i,g}}{(x_{i,g}-x_{k,g})^4}(0)-\frac{6|L_{ik}|^2\dot{x}_{i,L}}{(x_{i,L}-x_{k,L})^4}(0)+ \frac{2}{(x_{i,L}-x_{k,L})^3}\frac{d|L_{ik}|^2}{dt}(0)\right)+\mathcal{O}(t^4) \nonumber\\
 &=&-\frac{t^3}{3} \sum_{k\neq i}\frac{1}{(x_{i,L}-x_{k,L})^3}\frac{d|L_{ik}|^2}{dt}(0)+\mathcal{O}(t^4). \label{xgxL}
\end{eqnarray}

\begin{figure}[t]
	\centering
\includegraphics[width=18cm]{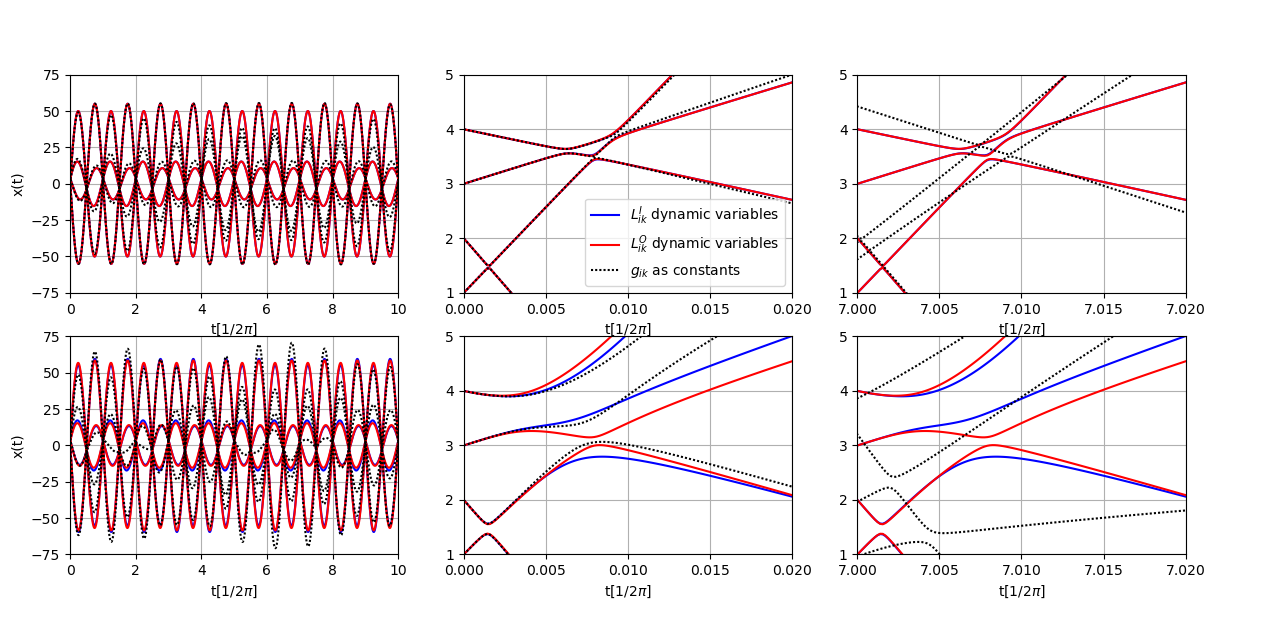} 
\caption{Positions of $N=4$ Calogero-Moser particles for constant $g_{ij}$ couplings, compared to $L^O$ and $L^I$ type variables, $|g_{ij}|=|L^{O,I}_{ij}(0)|$. The initial positions and momenta are $\bar{x}=(1,2,3,4)$ and $\bar{p}=(50,-50,15,-10)$. In the upper row  $\bar{g}=(g_{12},g_{13},g_{14},g_{23},g_{24},g_{34})=(1,2,3,1,2,1)$, that is $|g_{ij}|\approx 0.1\Delta x_{ij}\Delta p_{ij}$ (weak repulsion), and in the lower row $\bar{g}=10\cdot(1,2,3,1,2,1)$,that is $|g_{ij}|\approx\Delta x_{ij}\Delta p_{ij}$ (moderate repulsion). The first column shows the long timescale evolution: the dynamic $L$ variables enforce periodic motion, while the system with coupling constants deviates from periodicity. The middle column presents the first set of collisions. There is insignificant difference between constant, $L^I$ and $L^O$ couplings, when their absolute values are small in comparison to $\Delta x_{ij}\Delta p_{ij}$, a and a visible difference between the three when  $|g_{ij}|=|L_{ij}(0)|$ increase. The last column presents the collisions after the time of 7 periods: the difference between constants and matrix variables is prominent for both weak and strong repulsion. The legend applies to all plots.}
	\label{fig:N4positions}%
\end{figure}

Once we have expanded the expression for the positions, the result for the momenta is straightforward:
\begin{eqnarray} \label{pshort}
(p_{i,g}-p_{i,L})(t)&=&(p_{i,g}-p_{i,L})(0)+t(\dot{p}_{i,g}-\dot{p}_{i,L})(0)+\frac{t^2}{2}(\ddot{p}_{i,g}-\ddot{p}_{i,L})(0)+\mathcal{O}(t^3)= \nonumber \\
&=& -t^2 \sum_{k\neq i}\frac{1}{(x_{i,L}-x_{k,L})^3}\frac{d|L_{ik}|^2}{dt}(0)+\mathcal{O}(t^3).
\end{eqnarray} 
For the linear $X(t)=X_0+tY_0$ system the particles interact for a short time, and then scatter like almost free particles. Therefore to see the long term influence of the additional degrees of freedom, one can look at the system enclosed in a harmonic trap (\ref{CL}). Figure \ref{fig:N4positions} presents the short and long timescale motion of four particles with fixed initial positions and momenta, and different types of CM interactions (constant $g_{ij}$ values or dynamical $L_{ij}(t)$ with $|L_{ij}|(0)=g_{ij}$ but different initial phases). In the case of $|L_{ij}|$ being of order of typical relative position times typical relative momentum $\Delta x_{ij}\Delta p_{ij}$, the impact of the additional degree of freedom is visible in the short timescale. For weaker interactions (in comparison to $\Delta x_{ij}\Delta p_{ij}$) the difference is visible in the long timescale behaviour.


\subsection{The orthogonal vs. unitary setting}\label{ort}
As it was mentioned in Section~\ref{un-orth}, the real and symmetric subset of initial conditions $(X_0,Y_0)$ leads to a special subclass of trajectories confined to the subspace of real symmetric matrices $(X,Y)$, that can be treated independently as a Hamiltonian dynamics on the smaller phase space.  

We will concentrate here on the first interpretation, and denote the set of real, antisymmetric matrices $L$ by $L^O$. A complementary possibility corresponds to $L$ which is purely imaginary and symmetric with zeroes on the diagonal. The two subspaces are orthogonal in the sense of Hilbert-Schmidt product: $\langle L^O,L^I\rangle_{HS}=\tr (L^OL^I)=0$, since $L^O$ are antisymmetric and $L^I$ symmetric

In the unitary case we can use the decomposition $L_{ij}=L_{ij}^O+L_{ij}^I$, which upon substituting to (\ref{pb1}) gives 
\begin{eqnarray}
\{L^O_{ij},L^O_{kl}\}&=&\frac{1}{2}\left(\delta_{kj}L^O_{il}-\delta_{il}L^O_{kj}-\delta_{jl}L^O_{ik}+\delta_{ik}L^O_{lj}\right), \label{LPoissOO} \\
\{L^O_{ij},L^I_{kl}\}&=&\frac{1}{2}\left(\delta_{kj}L^I_{il}-\delta_{il}L^I_{kj}+\delta_{jl}L^I_{ik}-\delta_{ik}L^I_{lj}\right), \label{LPoissOI} \\
\{L^I_{ij},L^I_{kl}\}&=&\frac{1}{2}\left(\delta_{kj}L^O_{il}+\delta_{il}L^O_{jk}+\delta_{jl}L^O_{ik}+\delta_{ik}L^O_{lj}\right). \label{LPoissII}
\end{eqnarray}
The Poisson brackets (\ref{LPoissOO}) coincide, as they should, with (\ref{pbso}).

The most interesting difference between the orthogonal and unitary cases is the first time derivative of absolute values of the matrix elements. It can be directly checked from (\ref{Ldot}) that
\begin{equation}
\frac{d}{dt}|L_{ij}|^2 = 2\Re{\left[ L_{ij}\sum_{k\neq i,j}L_{jk}L_{ki}\left(\frac{1}{x_{ik}^2}-\frac{1}{x_{jk}^2} \right)\right] },
\end{equation}
and therefore:
\begin{equation}
\forall_{L^I} \frac{d|L_{ij}^I|}{dt} = 0,\hspace{5mm}\forall_{L^O} \frac{d|L_{ij}^O|}{dt}=4 L_{ij}^O\sum_{k\neq i,j}L_{jk}^OL_{ki}^O \left(\frac{1}{x_{ik}^2}-\frac{1}{x_{jk}^2} \right).
\end{equation}
In case of $L^I$ matrices the influence of $L$ variables on the positions (\ref{xgxL}) is of order of $t^4$, since the terms with  $\frac{d|L_{ik}|^2}{dt}(0)$ vanish and the first significant terms are a linear combination of $\frac{d^2|L_{ik}|^2}{dt^2}(0)$.

The special initial condition $L_0$ found in (\ref{init}) is of course of $L^I$ type, but it turns out that all imaginary $L$ matrices give rise to coupling which does not change at least for a short time. In the orthogonal case there is no stationary point in $L$ space, the couplings $|L_{ij}|$ between particles change immediately. 

\section{Reachable sets of the phase space}\label{sec:reach}
The set of points in a phase space accessible from given initial positions is important in many problems of control theory \cite{jurdjevich97}. In what follows we will concentrate on reachable matrices $L$. Let us start with the orthogonal settings. 

\subsection{Reachable $L$ in the orthogonal setting}

Since diagonal entries of real antisymmetric matrices vanish, we can parametrise every $L^O$ matrix with an $\mathbb{R}^{\binom{N}{2}}$ vector $\bar{l}$:
	\begin{equation}
	(L_{g})_{ij}=sign(j-i)l_{ij},\hspace{5mm} \bar{l}=(l_{12},l_{13},...,l_{N-1,N})
	\end{equation}
According to (\ref{intmotion}), the quantity  $\tr (L^2)=-2|\bar{l}|^2$ is an integral of motion. In other words, it means that during the evolution, $L_{\bar{l}^{\prime}}=OL_{\bar{l}}O^T$, the vector $\bar{l}$ undergoes an orthogonal 
rotation, $\bar{l}^{\prime}=M(O)\bar{l}$, with  $M(O)\in O\left(\binom{N}{2}\right)$, 
where we used notation explicitly exhibiting connection between $\bar{l}$ and 
$L_{\bar{l}}$ as well as between and $O$ and $M(O)$. 
%

Moreover, since
\begin{equation}
	L_{M(OO^{\prime})\bar{l}}= (OO^{\prime})L_{\bar{l}}(OO^{\prime})^T  
	= O O^{\prime}L_{\bar{l}}O^{\prime T} O^T
	= OL_{M(O^\prime)\bar{l}}O^T 
	= L_{M(O)M(O^\prime)\bar{l}},
\end{equation} 
$M$ is a group homomorphism, $M(OO^\prime) = M(O)M(O^\prime) $  of $O(N)$ and $O(\binom{N}{2})$.



In the special case of $N=3=\binom{3}{2}$ this is actually a group isomorphism given by 
\begin{equation}
M_3(O)=\left(\begin{array}{ccc} 0 & 0 & 1 \\
1 & 0      & 0 \\
0 & -1 & 0 \end{array} \right)O\left(\begin{array}{ccc} 0 & 1 & 0  \\
0 & 0      & -1 \\
1 & 0 & 0 \end{array} \right) \label{MO3}
\end{equation}
and it means that starting with any $L(0)=L_{\bar{l}}$ we can access an arbitrary $L(t)=L_{{\bar{l}}^{\prime}}$ by a correct choice of the rest of initial conditions. Having chosen ${\bar{l}}$ and ${\bar{l}}^{\prime}$, we find the orthogonal matrix $M$ which connects them: $M{\bar{l}}={\bar{l}}^{\prime}$, we reverse the relation (\ref{MO3}) to find the corresponding $O=O(t)$ matrix, and then ask for initial $X_{0,ii},Y_{0,ii}$ such that $X(t)$ is orthogonal in the basis of $O(t)$.
This is clearly not the case for $N>3$, where $M(O(N))=\{M(O): O\in O(N)\}$ is just a subgroup of $O(\binom{N}{2})$, which means that starting with a fixed $L(0)=L$ and varying freely the initial $X_{0,ii},Y_{0,ii}$, we can reach only a subset of all $L$ matrices.
This subset becomes easier to imagine, when we express both $O$ (limited to $SO(N)$ for simplicity) and $M(O)$ in their canonical forms. Let $N=2n+p$, where $p$ is either $0$ or $1$:
\begin{eqnarray}
R(\theta)&=&\left( \begin{array}{c c} \cos\theta & -\sin\theta \\
                                      \sin\theta & \cos\theta \end{array}\right),\hspace{5mm}  O_{2n+p}=\left( \begin{array}{c c c c} R(\theta_1) & 0 & ... & 0 \\
                                         0  & R(\theta_2) & ... & 0 \\
                                         ... & ... & ... & ... \\
                                         0  &  ... &  R(\theta_n) & 0 \\
                                         0  &  ... &      0       & \mathbb{1}_p \end{array}\right),\\  
M(O_{2n+p})&=&\left( \begin{array}{c c c c c} R(\theta_1-\theta_2) & 0 & ... & 0 & 0\\
0  & R(\theta_1+\theta_2) & ... & 0 & 0\\
... & ... & ... & ... \\
0  &  ... &  R(\theta_{n-1}-\theta_{n}) & 0 & 0\\
0  &  ... &    0  &  R(\theta_{n-1}+\theta_{n}) & 0 \\
0  &  ... &      0       & 0 & \mathbb{1}_{n(2p+1)} \end{array}\right).   \label{Mcanonical}                                
\end{eqnarray}
The rotations of a ${\bar{l}}$ vector are therefore limited to those which are parametrised by $n=(N-p)/2=\left[ N/2 \right]$ angles while the most general $SO(\binom{N}{2})$ matrix is parametrised with $\left[ n(N-1)/2\right]\geq n$ independent angles. The equality occures for $N=3$, in which case the above canonical forms are identical. This gives limitations on the $L(t)=L_{{\bar{l}}^\prime}$ matrices accessible from a given $L(0)=L_{\bar{l}}$. Having two vectors, ${\bar{l}},{\bar{l}}^\prime\in \mathbb{R}^{(N^2-N)/2}$, we need to find the canonical form of the orthogonal matrix connecting them and check if it can be expressed whith at most $\left[ N/2 \right]$ angles, as in (\ref{Mcanonical}). If it is possible, the corresponding $L(t)$ is accessible from $L(0)$ with the correct choice of initial positions and momenta $X_{0,ii},Y_{0,ii}$. Otherwise, $L(t)$ lies outside the so called reachable set of $L(0)$.

\subsection{The image of $L(t)=U(t)L(0)U^{\dagger}(t)$}

In the previous section we have discussed the question what is the subset of $L$ matrices that could be accessed \textit{via} the generalized CM dynamics from a given initial $L_0$ in the orthogonal setting. Here we shall extend the discussion to the general, unitary case.

In the orthogonal setting, the $L$ matrices spanned the full $\mathfrak{so}N$ algebra, and the action of the $SO(N)$ group on this algebra of course preserves it. In the general, unitary setting $L$ matrices do not span the $\mathfrak{su}N$, but its subspace with vanishing diagonals, and the action of the $SU(N)$ group does not preserve this subspace. Therefore the matrix group approach used in the previous section will no longer be useful. 

The reachable set can be found \textit{via} the characterictic polynomial of $L_0$ (or $\pm iL_0$ if we wish to deal with a Hermitian matrix). The unitary evolution of course preserves the eigenvalues, and therefore must preserve the coefficients of the characteristic equation. But as mentioned before, $U(t)$ must also preserve the vanishing diagonal and this, in turn, gives more limitations on $L_{ij}(t)$. Let us demonstrate this with the simplest example of $N=3$. The most general form of $L(t)$ is the following:
\begin{equation}
L(t)=i\left(\begin{array}{ccc} 0 & l_{12}(t)e^{i\phi_{12}(t)} & l_{31}(t)e^{-i\phi_{31}(t)} \\
                         l_{12}(t)e^{-i\phi_{12}(t)}  & 0     & l_{23}(t)e^{i\phi_{23}(t)}  \\
                         l_{31}(t)e^{i\phi_{31}(t)}   & l_{23}(t)e^{-i\phi_{23}(t)} &  0    \end{array}\right),
                 \hspace{5mm}
L(0)=i\left(\begin{array}{ccc} 0 & l_{12}e^{i\phi_{12}} & l_{31}e^{-i\phi_{31}} \\
l_{12}e^{-i\phi_{12}}  & 0     & l_{23}e^{i\phi_{23}}  \\
l_{31}e^{i\phi_{31}}   & l_{23}e^{-i\phi_{23}} &  0    \end{array}\right),              
\end{equation}  
where $l_{ij}=|L_{ij}|\geq 0$ and $\Phi_{123}(t)=\phi_{12}(t)+\phi_{23}(t)+\phi_{31}(t)\in\left[-\pi/2,\pi/2\right]$ (this is a sufficient set due to gauge invariance (\ref{ginv})). The characteristic equation at time $t$ reads:
\begin{equation}
\lambda^3-|{\bar{l}}(t)|^2\lambda +2\cos(\Phi_{123}(t))l_{12}(t)l_{23}(t)l_{31}(t)=0,
\end{equation}
where ${\bar{l}}(t)=(l_{12},l_{23},l_{31})(t)$. The constrains for the time dependent $l_{ij}$ and $\phi_{ij}$ functions stem from the initial values:
\begin{equation}
|{\bar{l}}(t)|=|{\bar{l}}|, \hspace{5mm} \cos(\Phi_{123}(t))l_{12}(t)l_{23}(t)l_{31}(t)=\cos(\Phi_{123})l_{12}l_{23}l_{31}.
\end{equation}
The first equation states that ${\bar{l}}(t)$ is confined to a sphere of the radius given at $t=0$ (and it is equivalent to the conservation of $\tr L^2$). In the orthogonal case $\Phi_{123}=\pm \pi/2$ and the first equation is the only constraint. This is another proof of the transitive action of $O(3)$ on $3\times 3$ L matrices with constant $Tr(L^2)$ presented in section (\ref{ort}).

The second equation says that ${\bar{l}}(t)$ is a point on a paraboloid $x\cdot y\cdot z = p_0$, from a family of paraboloids given by $p_0=\cos(\Phi_{123})l_{12}l_{23}l_{31}/\cos(\Phi_{123})(t)\geq \cos(\Phi_{123})l_{12}l_{23}l_{31}$. The two equations limit ${\bar{l}}(t)$ to those paraboloids which intersect with the sphere. They cut out a spherical cap centered at the point $x=y=z=\frac{|{\bar{l}}|}{\sqrt{3}}$, and the edge of the cap is the circular intersection of the sphere and $x\cdot y\cdot z = \cos(\Phi_{123})l_{12}l_{23}l_{31}$. The closer $\cos(\Phi_{123})$ is to unity (and $\Phi_{123}$ - to zero), the smaller is the cap. In particular for $l_{12}=l_{23}=l_{31}=\frac{g}{\sqrt{3}}$ and $\Phi_{123}=0$, that is for the stationary $L_0$ matrix found in section (\ref{init}), it shrinks to a point. From this point of view, $L_0$ is stationary because there are no other points with the same eigenvalues.  

\begin{figure}[h]
	\centering
	\subfigure[$\Phi_{123}=\pi/2$]{\includegraphics[width=0.23\textwidth]{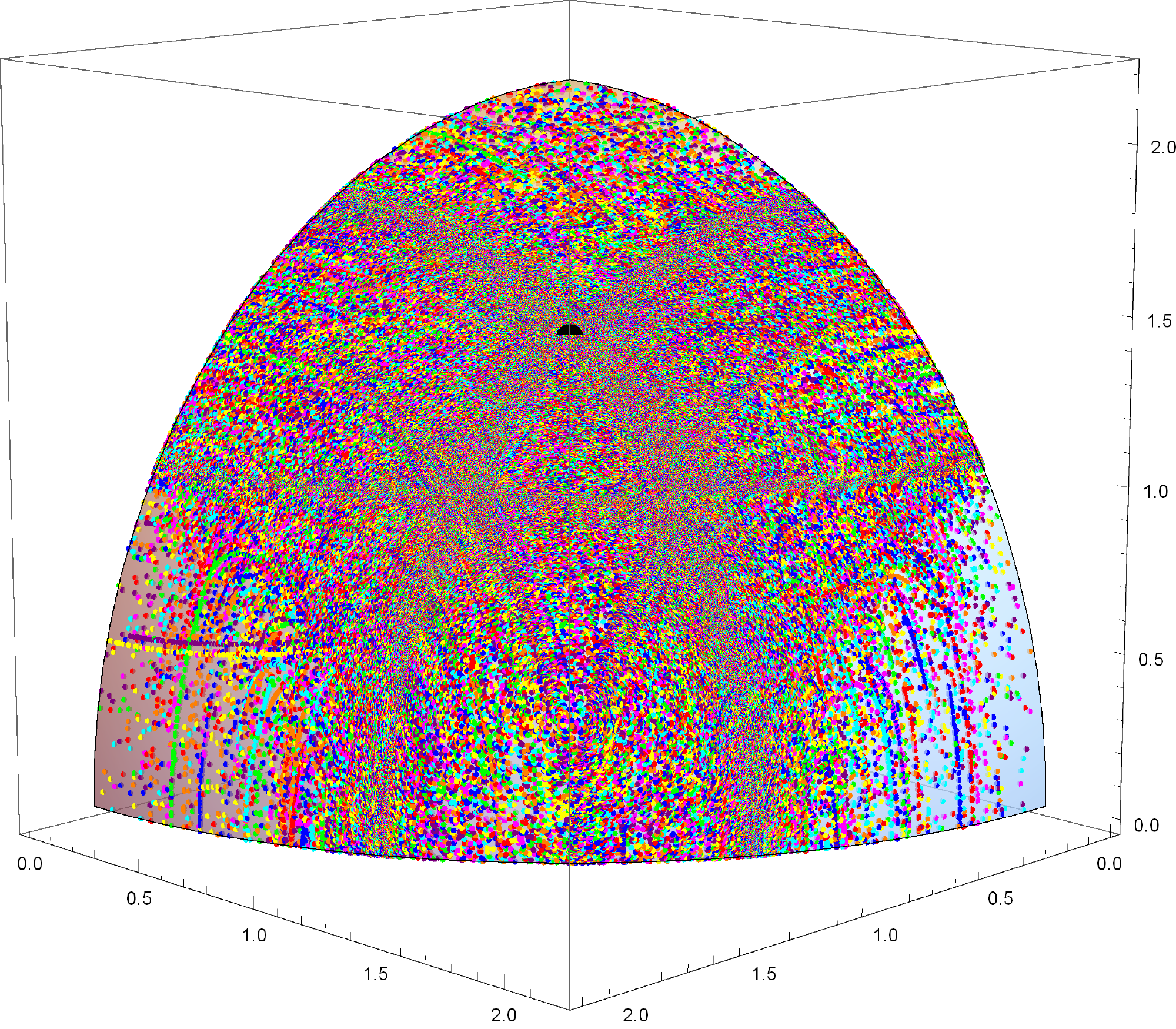}} 
	\subfigure[$\Phi_{123}=\pi/3$]{\includegraphics[width=0.23\textwidth]{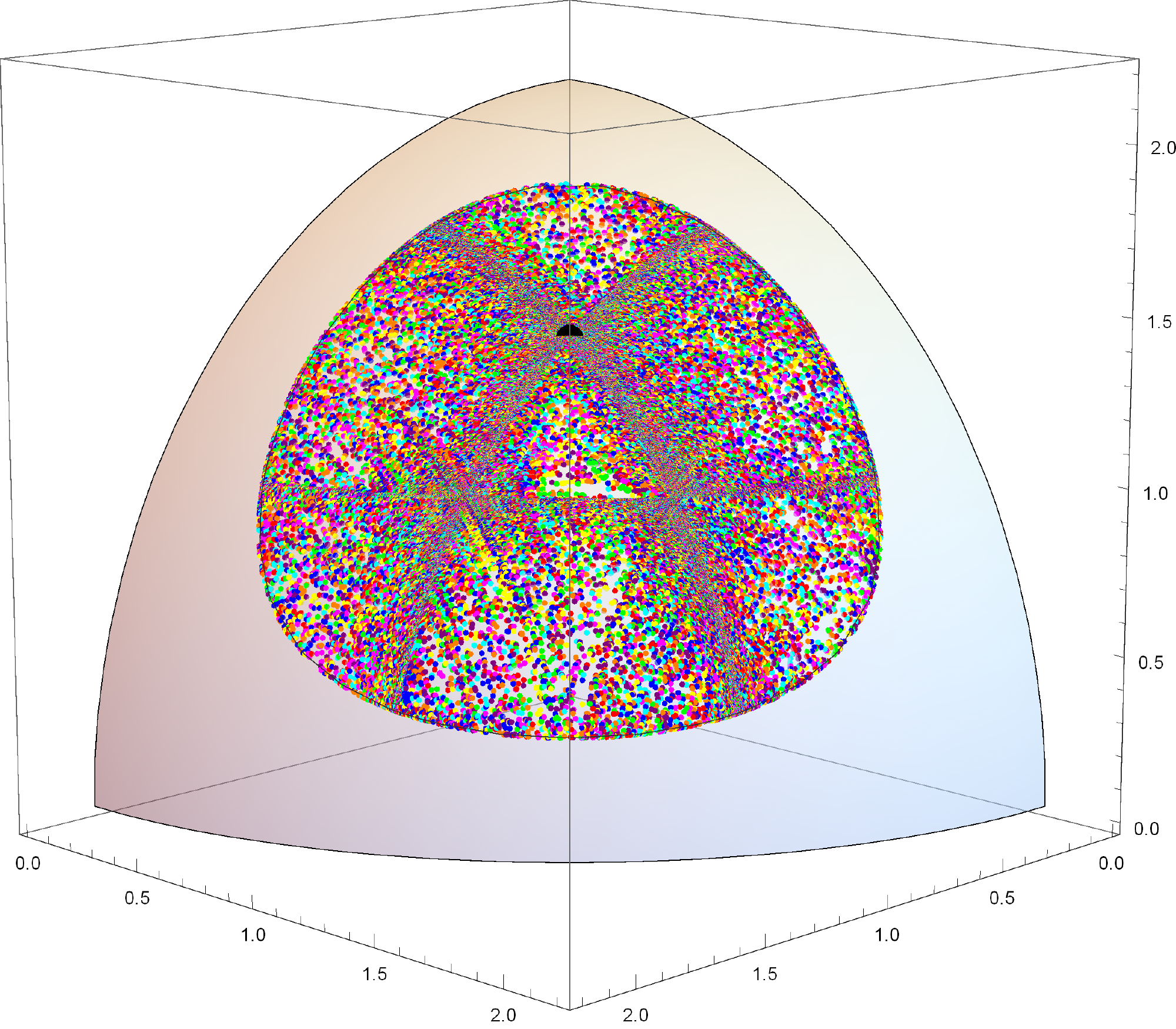}}
	\subfigure[$\Phi_{123}=\pi/6$]{\includegraphics[width=0.23\textwidth]{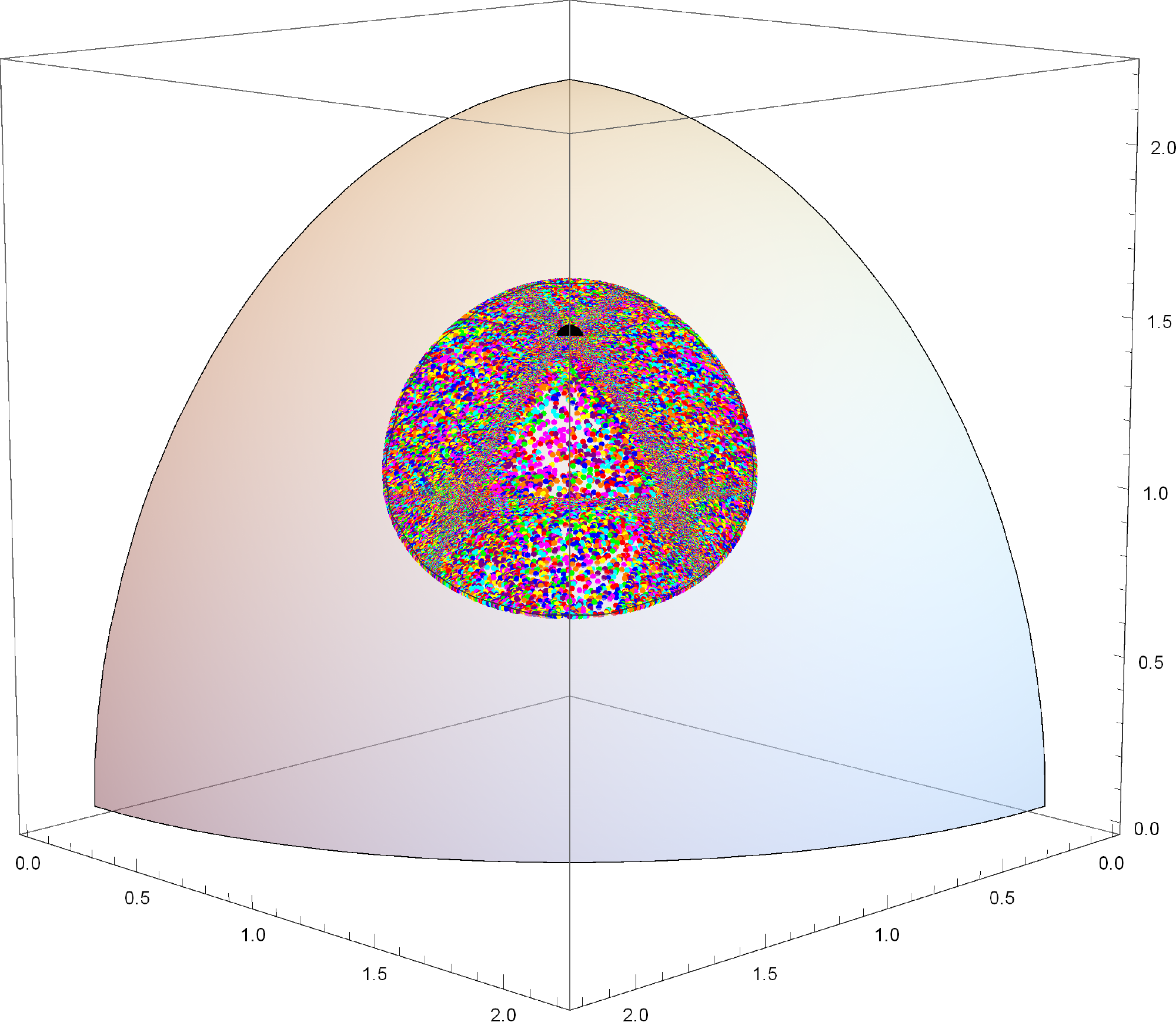}} 
	\subfigure[$\Phi_{123}=0$]{\includegraphics[width=0.23\textwidth]{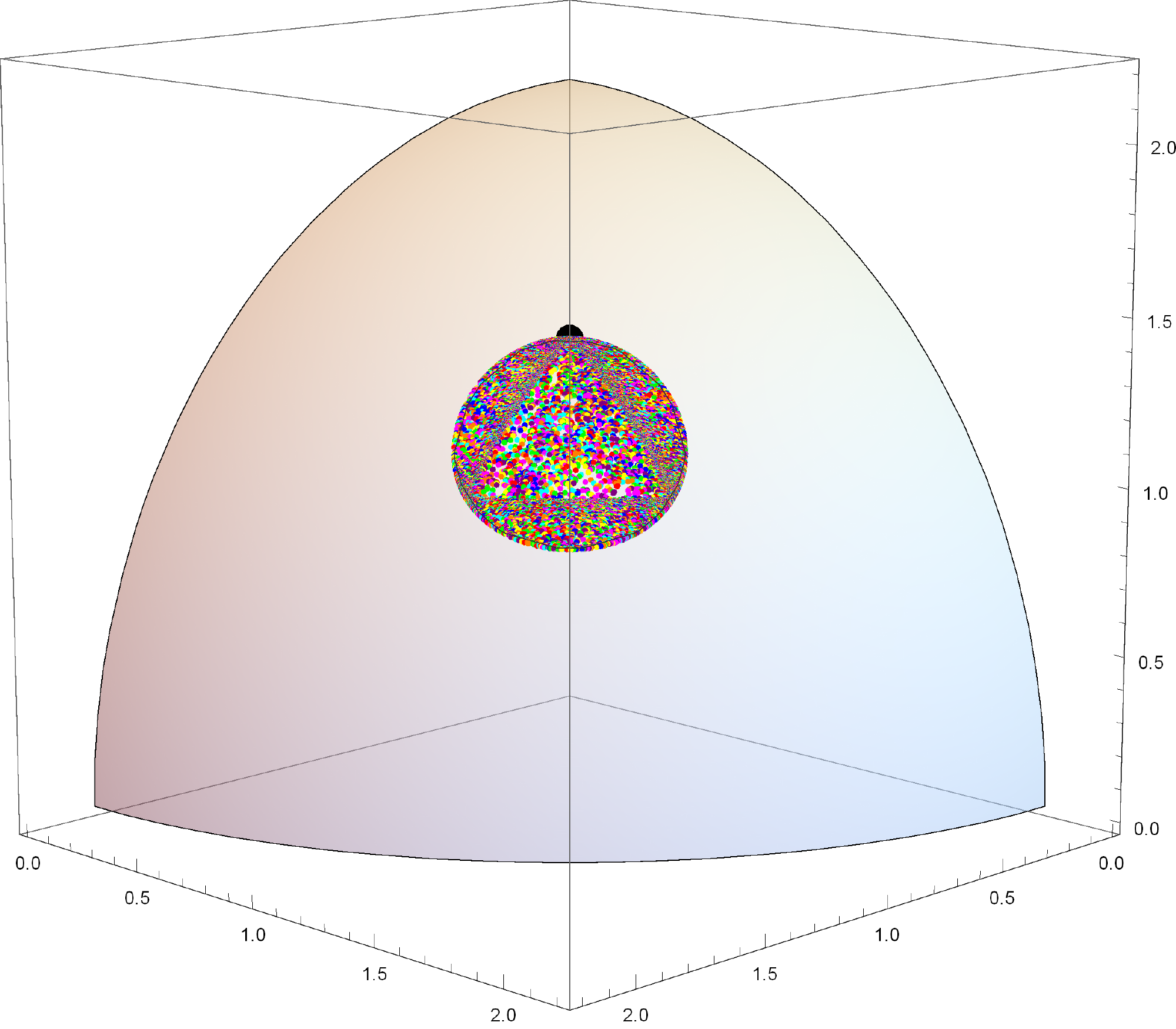}}
	\caption{Regions accessed by $L(t)$ for the initial condition ${\bar{l}}=(1,1,\sqrt{2})$ (black dots) and various values of $\Phi_{123}$. Each image consists of 5000 trajectories generated from random initial positions and momenta. The numerical results coincide with the analytical predictions and serve as an illustration.}
	\label{fig:images}
\end{figure}

\section{Vectorial degrees of freedom}\label{sec:vectorial}
One can construct generalized CM systems in which the internal degrees of freedom are more prominent than the $L_{ij}$ variables discussed until now \cite{gibbons84}, \cite{krichever95}. In particular, they are strictly one-particle properties, rather than (seemingly) two-body features described by the $L_{ij}$ variables. In what follows we will, however, show that these one-particle variables can reconstruct every $L(t)$ trajectory defined by the matrix model described in section \ref{matrix_dyn}. Next, we shall use this description to explore the problem of $|L(t)|=|L(0)|$ (that is ordinary CM) trajectories posed in section \ref{init}.

We start with an extended phase space parameterized by `canonical variables' $(x_i,p_i)$ of $N$ particles on a line and assigning vectorial degrees of freedom to each particle. To this end we extend the initial $(x_i,p_i)$ phase space  by $|e_i)\in \mathbb{C}^d$, and their duals $(f_i|$, $i=1,2,...,N$ and define a symplectic structure by the following symplectic form,
\begin{equation}
\omega =\sum_{i=1}^N dx_i\wedge dp_i +i\cdot d\cdot (f|\wedge d|e) \label{omegaEF}
\end{equation}
The following Hamilton function on this phase space:
\begin{equation}
\mathcal{H}=\frac{1}{2}\sum_{i} p_i^2 + \frac{1}{2}\sum_{i\neq j}\frac{(f_i|e_j)(f_j|e_i)}{(x_i-x_j)^2} \label{CMef}
\end{equation}
is another possible generalization of the ordinary CM system. 

\subsection{The relation between the vectorial and matrix formulation}\label{VecCom}
We need to understand the relation between the dynamics governed by (\ref{CML}) and (\ref{CMef}). The corresponding equations of motion are the following:
\begin{eqnarray}
\dot{x}_i&=& \frac{\partial H}{\partial p_i} = p_i, \\
\dot{p}_i&=& -\frac{\partial H}{\partial x_i}=\sum_{k\neq i} \frac{2(f_i|e_k)(f_k|e_i)}{(x_i-x_k)^3}, \\
\frac{d|e_i)}{dt}&=&-i\cdot\frac{\partial H}{\partial (f_i|}=-i\sum_{k\neq i}\frac{|e_k)(f_k|}{(x_i-x_k)^2}|e_i), \label{ei} \\
\frac{d(f_i|}{dt}&=&i\cdot\frac{\partial H}{\partial |e_i)}=i(f_i|\sum_{k\neq i}\frac{|e_k)(f_k|}{(x_i-x_k)^2}. \label{fi}
\end{eqnarray} 
There are two useful conclusions from (\ref{ei}) and (\ref{fi}). Firstly, the  scalar products $(f_i|e_i)$ are constant in time. Secondly, the dimension of the $(e,f)$ vector spaces is effectively $d\leq N$ \cite{gibbons84}. This stems from the fact that $|e_i(t))\in span\{|e_{0,i})\}$, and as there are N vectors, at most N of which linearily independent, the space spanned by the evolving vectors is at most N-dimensional (and fixed by the vectors given at t=0). 
Comparing the above to (\ref{xdot})-(\ref{Ldot}) we observe that $i\cdot(f_i|e_j)$  obey an equation of motion which is similar to (\ref{Ldot}) obeyed by $L_{ij}$. The time derivative calculated with the use of (\ref{ei}) and (\ref{fi}) is
\begin{equation}
\frac{d}{dt}\left(i\cdot(f_i|e_j)\right)= \sum_{k \neq i,j} i(f_i|e_k)\cdot i(f_k|e_j)\left(\frac{1}{(x_i-x_k)^2}-\frac{1}{(x_j-x_k)^2}\right)+\frac{1}{x_{ij}^2}(f_i|e_j)\cdot ((f_j|e_j)-(f_i|e_i)), \label{efDot}
\end{equation}
which differs from (\ref{Ldot}) by a term which vanishes if the vectors are equally normalised $(f_i|e_i)=g$. Moreover, the Poisson brackets obtained with the use of $e,f$ part of the symplectic form (\ref{omegaEF})
\begin{equation}
\{i(f_i|e_j),i(f_k|e_l)\}=i\left(\sum_m \frac{\partial (f_i|e_j)}{|e_m)}\frac{\partial (f_k|e_l)}{(f_m|}-\frac{\partial (f_k|e_l)}{|e_m)}\frac{\partial (f_i|e_j)}{(f_m|}\right) = i\left(\delta_{jk}(f_i|e_l)-\delta_{il}(f_k|e_j)\right),
\end{equation} 
coincide with $\left\{ L_{ij},L_{kl}\right\} =\delta_{jk}L_{il}-\delta_{il}L_{kj}$ (\ref{pb1}).
This means that once we find such initial $|e_{0,i}), (f_{0,i}|$  which recover a given anti-Hermitian $L_0$ matrix via $L_{0,ij}=i\cdot(f_{i,0}|e_{j,0})$, and $(f_{i,0}|e_{i,0})=g$, the two flows will be equivalent. The set of such initial conditions is therefore restricted by the hermicity condition:
\begin{equation}
\forall_{i,j}(f_i|e_j)=(f_j|e_i)^*\label{hermitEF}
\end{equation}
which is also sufficient for (\ref{CMef}) to be real.The necessary condition would be $(f_i|e_j)=\pm(f_j|e_i)^*$, where the minus sign means an attractive interaction which is outside the scope of this study. Expressing the above with rectangular matrices
\begin{equation}
E =\left(|e_1) |e_2) ... |e_N)\right), \hspace{5mm}
F =\left(|f_1) |f_2) ... |f_N)\right)^\dagger, \label{defFE}
\end{equation}
any Hermitian $FE$ matrix with a constant diagonal defines a repulsive (\ref{CMef}) system. Once the initial conditions coincide, $L_{ij}(0)=i(FE)_{ij}(0)$, and the diagonal elements are all equal $(FE)_{ii}(0)=g$, the trajectories will coincide as well: $L_{ij}(t)=i(FE)_{ij}(t)$. Now two questions need to be answered: 1. does every $L$ matrix decompose into $|e),(f|$ vectors, 2. does every vectorial $|e),(f|$ model translate to an $L$ formulation? In other words: do (\ref{CML}) and (\ref{CMef}) coincide, overlap or one contains the other?

\subsubsection{Decomposition of $L$ into vectorial form}
First let us decompose an $L$ matrix into $|e_i),(f_i|$ vectors. We know that every Hermitian, positive definite $N\times N$ matrix $M$ can be expressed with the so called Cholesky decomposition $M=\mathcal{E}^{\dagger}\mathcal{E}$, where $\mathcal{E}$ is upper (or lower, depending on the convention) triangular, and its column vectors $\left\lbrace |\epsilon_i)\right\rbrace$ span the full $N$-dimensional space. A positive semidefinite matrix can be decomposed like this as well, only the column vectors of $\mathcal{E}$ will span a subspace of dimension $N-\mu^0$, where $\mu^0$ is the multiplicity of the $0$ eigenvalue. Negative (semi) definite matrices will be simply $M=-\mathcal{E}^{\dagger}\mathcal{E}$. The $L$ matrices are antihermitian and non-definite, yet we can easily adjust such a matrix to a decomposable form
\begin{equation}
\pm iL+\mathbb{1}g=\mathcal{E}_g^{\dagger}\mathcal{E}_g,
\end{equation}
with a large enough value of $g$, so to make it positive semi-definite. This adjustment is of course not unique, but throughout this study we choose the one which minimizes the dimension of $span(|\epsilon_i))$. The extreme eigenvalues $i|\lambda_+|$ and $-i|\lambda_-|$ of $L$ give us two positive semidefinite matrices to choose from:\begin{equation}
\pm iL+\mathbb{1}|\lambda_\pm|=\mathcal{E}_{\pm}^{\dagger}\mathcal{E}_{\pm}.
\end{equation}
The one with the higher multiplicity results in the smallest possible subspace spanned by the column vectors $|\epsilon_i)$ and will be denoted as $\mathcal{E}$. The column vectors of $\mathcal{E}$ give a valid expression of the initial conditions given by $L$ 
\begin{equation}
L_{ij}=i(\epsilon_i|\epsilon_j),\hspace{5mm} (\epsilon_i|\epsilon_i)=|\lambda_{\pm}|
\end{equation}
in terms of $|e_i)=|f_i)=|\epsilon_i)$. The symmetry of the (\ref{CMef}) and resulting equations of motion (\ref{ei}),(\ref{fi}) implies that having found the $\left\lbrace |\epsilon_i)\right\rbrace $ decomposition of $L$, we automatically obtain a vast set of other choices:
\begin{equation}
|e_i)=W|\epsilon_i),\hspace{5mm}(f_i|=(\epsilon_i|W^{-1},\hspace{5mm} W\in GL(N,\mathbb{C})
\end{equation}

We can conclude that every valid $L$ matrix can be expressed with a single set of $N$ vectors of equal length. 

\subsubsection{Translating Hermitian $FE$ into $L$ matrices}
We know that every antihermitian matrix with a vanishing diagonal can be expressed with $N$ complex vectors of equal length. Yet the vectorial formulation gives us freedom to set the initial conditions with $2N$ vectors $|e_i),|f_i)\in\mathbb{C}^d$ and the only constraint is that $FE$ (\ref{defFE}) is Hermitian. We need to check if this broader choice can give us anything more that the $L$ matrix formulation. We already know that if $(FE)_{ii}=g$, we can define the initial $L$ as:
\begin{equation}
L=i(FE-\mathbb{1}g),
\end{equation}
and according to (\ref{Ldot}) and (\ref{efDot}), the corresponding flows of $L_{ij}$ and $i(f_i|e_j)$ will coincide. What we need to check is the role of the nonvanishing $(f_i|e_i)-(f_j|e_j)$ term. If it can be removed by a choice of gauge, that is if there exist $\phi_i(t),i=1,2,...,N$ such that all $i(f_i|e_j)e^{i(\phi_i-\phi_j)}$ obey (\ref{Ldot}), then the $FE$ matrices with different diagonal elements are equivalent to the $L$ formulation as well. Let us define $\mathcal{L}_{ij}=i(f_i|e_j),(f_i|e_i)=c_i$,where the values of $c_i$ are constants of motion. Let us write the time derivative of $\mathcal{L}_{ij}e^{i(\phi_i-\phi_j)}$ and find the equations for the suitable gauge:
	
\begin{eqnarray}
\frac{d}{dt}\left(\mathcal{L}_{ij}e^{i(\phi_i-\phi_j)}\right) -\dot{\mathcal{L}}_{ij}e^{i(\phi_i-\phi_j)}&=&\mathcal{L}_{ij}e^{i(\phi_i-\phi_j)}i(\dot{\phi_i}-\dot{\phi_j})=\frac{i}{x_{ij}^2}(c_j-c_i)\mathcal{L}_{ij}e^{i(\phi_i-\phi_j)}, \\
\dot{\phi_i}-\dot{\phi_j} &=& \frac{1}{x_{ij}^2}(c_j-c_i). \label{cijGauge}
\end{eqnarray} 
 For a set of functions $\phi_1(t),\phi_2(t),...,\phi_N(t)$ to exist, every triple $(k,l,m)$ of equations must sum up to 0 at every instant of time:
 \begin{equation}
 0=\dot{\phi_k}-\dot{\phi_l}+\dot{\phi_l}-\dot{\phi_m}+\dot{\phi_m}-\dot{\phi_k}=\frac{1}{x_{kl}^2}(c_l-c_k)+\frac{1}{x_{lm}^2}(c_m-c_l)+\frac{1}{x_{mk}^2}(c_k-c_m)
 \end{equation}

The numerator of the right hand side expanded in $x_{kl}=a,x_{lm}=b,x_{km}=a-b$ will have coefficients that vanish for any $a,b$ only if $c_k=c_l=c_m$. Therefore, as we are free to choose any initial positions $x_i(0)$, and momenta  the (\ref{cijGauge}) conditions have no solution. Of course we can tune the initial positions so that they are met at time $t=0$, but in $t=\delta t$ it will no longer be true. We conclude that $FE$ matrices with different diagonal elements do not translate into an $L$ matrix formulation. They define a flow of $i(f_i|e_j)$ matrix elements which is not gauge equivalent to any $L_{ij}$ evolution.

\subsubsection{Redundancy of the second set of vectors}
The last question which needs to be answered is why a single set of $|\epsilon_i)$ vectors is enough to formulate the vectorial model (\ref{CMef}). If the $FE$ matrix is positive semidefinite, then it can be decomposed as $\mathcal{E}^{\dagger}\mathcal{E}$ and the dimension of $span(|\epsilon_i))$ is equal to the dimension of the intersection of $span(|e_i))$ and $span(|f_i)$. Otherwise it can be made positive semidefinite by adding $\mathbb{1}\cdot g$, which has no influence on the equations of motion, and then decomposed. What is more, this is true regardless of the values of $(f_i|e_i)$. If they are all the same, the $FE$ matrix elements evolve as elements of an $L$ matrix. If they are different, they evolve on a different manifold, nevertheless there are only the differences $(f_i|e_i)-(f_j|e_j)$ present in the equations of motion, so shifting all of them has no impact on the behaviour of $FE$ matrix elements.

We can conclude that the vectorial $|e_i),(f_i|$ formulation is equivalent to the $L=[X,Y]$ formulation of the Calogero-Moser system if and only if the diagonal elements are all equal: $(f_i|e_i)=g$. In this case  both $L$ and $FE$ can be expressed via the Cholesky decomposition with a set of $N$ complex vectors of equal length. If the diagonal elements are different, the differences appear in the equations of motion and are imposiible to remove with a choice of gauge. This means, that the unequal values of $(f_i|e_i)$ take us beyond the $L$ matrix formulation.

This result shows that the $L_{ij}$ variables do not have a generically two-particle character - the interactions in the generalised CM model arrise as functions of one-particle observables. This is a valuable hint for the quantisation of the system.

\subsection{Orbits of L classified by rank}\label{rorbits}
The previous section showed that any $L$ can be expressed by a single set of complex vectors $\mathcal{E}=(|\epsilon_1) |\epsilon_2)...|\epsilon_N))$ of the same length. The equations of motion
\begin{equation}
\frac{d|\epsilon_i)}{dt}=-i\sum_{k\neq i}\frac{|\epsilon_k)(\epsilon_k|}{x_{ik}^2}|\epsilon_i)=-i\cdot M_i(t)|\epsilon_i), \label{epsilont}
\end{equation} 
where $x_{ik}=x_i-x_k$, imply that $r=rank(\mathcal{E})$ is invariant, because the evolving vectors stay in the span of the initial ones. The aim of this section is to establish the classification of orbits of $L$ via the value of r. The N vectors belong to an r-dimensional subspace of $\mathbb{C}^N$. Therefore we can choose a basis in $\mathbb{C}^N$ in which at most first r components of $\{|\epsilon_i)\}$ are nonzero, and treat them as elements of $\mathbb{C}^r$. The general solution of (\ref{epsilont}) is the following:
\begin{equation}
|\epsilon_i)(t)=U_i(t)|\epsilon_i)(0)
\end{equation}
where $U_i(t)=Texp\left(-i\int_0^t M_i(\tau)d\tau\right)\in U(r)$. The unitarity of $U_i(t)$ expresses the fact that $(\epsilon_i|\epsilon_i)$ are constants of motion. These matrices are of course very sensitive to the initial positions $x_i(0)$ and momenta $p_i(0)$, but we want to know the properties of $|\epsilon_i)$ variables independent of the initial conditions other that $|\epsilon_i)(0)$. In particular we shall answer the question posed in (\ref{init}), whether  $L_0=ig(|e\rangle\langle e|-\mathbb{1})$ resulting in an ordinary CM system is unique. 

\subsubsection{Vectorial gauge invariance}
Just as the $L$ matrices (\ref{ginv}), the vectorial variables also fall into equivalence classes due to gauge symmetry. In case of vectors this is the simplest possible $U(1)$ symmetry: $[|\epsilon_i)]=[|\epsilon_i')] \iff |\epsilon_i')=e^{i\phi}|\epsilon_i)$ and it translates automatically to gauge equivalent $L$ matrices formed as $L_{ij}=i(\epsilon_i|\epsilon_j)$ and $L'_{ij}=i(\epsilon_i|\epsilon_j)e^{i(\phi_j-\phi_i)}=L_{ij}e^{i(\phi_j-\phi_i)}$. In other words equivalent vectors define the same one-dimensional eigenspace of a projection operator $P_i=|\epsilon_i)(\epsilon_i|$ (we normalize the vectors to unity, which can be done without loss of generality).

\subsubsection{Rank r=1 as the ordinary CM system}
As mentioned before, the vectors evolve within the span of the initial vectors and their length is a constant of motion. Therefore, if we choose the initial $|\epsilon_i)(0)=e^{i\phi_i}|\epsilon)$, which corresponds to $r=1$, the evolving vectors will stay in this one dimensional subspace, their length is constant, and all that can change in time is the phases: $|\epsilon_i)(t)=e^{i\phi_i(t)}|\epsilon)$. Note that regardless of the phases and the choice of the $|\epsilon)$ vector this initial condition corresponds to 
\begin{equation}
L_{ij}(t)=i|\epsilon|^2\cdot e^{i(\phi_j(t)-\phi_i(t))}
\end{equation}
that is the special case of the ordinary CM system found in section \ref{init}. 

\subsubsection{First time derivative of $|(\epsilon_i|\epsilon_j)|^2$ for $r>1$}
For any rank $r>1$ there is room for nontrivial evolution of $|(\epsilon_i|\epsilon_j)|^2(t)$. A heuristic argument for this is that for $(\epsilon_i|\epsilon_j)(t)$ to differ from $(\epsilon_i|\epsilon_j)(0)$ only by a phase factor, the two vectors should aquire phase factors only (as in the $r=1$ case) or rotate in a strictly synchronised way, which is likely only for isolated positions and momenta. Let us now look for such sets of initial vectors $|\epsilon_i)(0)$, that result in stationary couplings:
\begin{equation}
\forall_{i\neq j}|(\epsilon_i|\epsilon_j)|^2(t)=|(\epsilon_i|\epsilon_j)|^2(0).
\end{equation}
All time derivatives at $t=0$ must vanish for a constant function, but the vanishing of first two is the necessary condition:
\begin{eqnarray}
\frac{d}{dt}|(\epsilon_i|\epsilon_j)|^2(0)&=&i\sum_{k\neq i,j} (\epsilon_i|[P_k,P_j]|\epsilon_i)\left(\frac{1}{x_{ik}^2}-\frac{1}{x_{jk}^2}\right)(0) =0 \label{ddt} \\
\frac{d^2}{dt^2}|(\epsilon_i|\epsilon_j)|^2(0)&=& i\sum_{k\neq i,j} \frac{d}{dt}\left[(\epsilon_i|[P_k,P_j]|\epsilon_i)\left(\frac{1}{x_{ik}^2}-\frac{1}{x_{jk}^2}\right)\right](0) = 0 \label{d2dt2}
\end{eqnarray} 
 First we search for vectors which satisfy (\ref{ddt}), regardless of the initial positions. The $\binom{N}{3}$ position dependent factors are expressed only by $N-1$ independent relative distances. This is why we cannot automatically assume that all $(\epsilon_i|[P_k,P_j]|\epsilon_i)$ must vanish. Yet it can be shown (\ref{appddt}) that when we expand (\ref{ddt}) for a fixed pair $i<j$ in terms of independent distances ( for example $x_{i1},x_{i2},...,x_{i,i-1},x_{i,i+1},...,x_{iN}$) over a common denominator, the obtained polynomial expression will be identically zero if and only if all the coefficients, that is $(\epsilon_i|[P_k,P_j]|\epsilon_i)$, vanish. This condition can be rewritten as:
\begin{equation}
\forall_{k,i,j} Im((\epsilon_i|\epsilon_k)(\epsilon_k|\epsilon_j)(\epsilon_j|\epsilon_i))=0 \label{Im0}
\end{equation}
There are only two types of initial conditions which satisfy this equality: either all vectors project on the same one-dimensional subspace: $|\epsilon_i)=e^{i\phi_i}|e)$ (this is of course the $r=1$ case discussed before) or all vectors are (gauge equivalent to) real $|\epsilon_i)=|e_i^R)$. It can be shown (\ref{appIm0}) that in case of $r\geq 2$ this is the only possibility to satisfy (\ref{Im0}). In other words $L^I$ type matrices (\ref{ort}) are the only ones which have a vanishing first time derivative. 

\subsubsection{Second time derivative of $|(\epsilon_i|\epsilon_j)|^2$ for colliding pairs}
Now we find the sets of real vectors which satisfy (\ref{d2dt2}) as well. The time derivative can be expanded as follows:
\begin{equation}
\frac{d^2}{dt^2}|(\epsilon_i|\epsilon_j)|^2(0)= i \sum_{k\neq i,j} \frac{d}{dt}\left[(\epsilon_i|[P_k,P_j]|\epsilon_i)\right]\left(\frac{1}{x_{ik}^2}-\frac{1}{x_{jk}^2}\right)+(\epsilon_i|[P_k,P_j]|\epsilon_i)\frac{d}{dt}\left(\frac{1}{x_{ik}^2}-\frac{1}{x_{jk}^2}\right). \label{d2dt22}
\end{equation}
In cases when (\ref{ddt}) is satisfied, the terms proportional to $(\epsilon_i|[P_k,P_j]|\epsilon_i)$ vanish and we are left with
\begin{equation}
 \sum_{k\neq i,j}\left(\frac{1}{x_{ik}^2}-\frac{1}{x_{jk}^2}\right) \frac{d}{dt}\left[(\epsilon_i|[P_k,P_j]|\epsilon_i)\right] =0. \label{d2dt23}
 \end{equation}
Further expansion of the derivative:
\begin{eqnarray}
& &\frac{d}{dt}\left[(\epsilon_i|[P_k,P_j]|\epsilon_i)\right]= (\dot{\epsilon}_i|[P_k,P_j]|\epsilon_i)+(\epsilon_i|[P_k,P_j]|\dot{\epsilon}_i)+(\epsilon_i|\frac{d}{dt}([P_k,P_j])|\epsilon_i)= \nonumber \\
& &= i\left(\sum_{l\neq i}\frac{1}{x_{il}^2}(\epsilon_i|[P_l,[P_k,P_j]]|\epsilon_i)+\sum_{l\neq j}\frac{1}{x_{jl}^2}(\epsilon_i|[P_k,[P_j,P_l]]|\epsilon_i)+\sum_{l\neq k}\frac{1}{x_{kl}^2}(\epsilon_i|[P_j,[P_l,P_k]]|\epsilon_i)\right) \label{d2dt24}
\end{eqnarray} 
makes use of the fact, that $\dot{P}_k=i\sum_{l\neq k}x_{kl}^{-2}[P_k,P_l]$. The last equality in (\ref{d2dt24}) used in (\ref{d2dt23}) gives us a double sum with many terms which are difficult to manage in general. Yet we require (\ref{d2dt23}) to be true for any initial distances $x_{ab}$, so we can choose for example a special configuration in which two particles, $i^{th}$ and $(i+1)^{st}$, are close to each other, while the others are much further away: $|x_{i,i+1}|<< |x_{kl}|$. This lets us take only the dominant term, proportional to $|x_{i,i+1}|^{-4}$ (others are of order $|x_{i,i+1}|^{-2}|x_{kl}|^{-2}$, or even smaller $|x_{kl}|^{-2}|x_{mn}|^{-2}$) and demand that it will be zero:
\begin{equation}
\frac{2|(\epsilon_i|\epsilon_{i+1})|^2}{x_{i,i+1}^4}\left(|(\epsilon_i|\epsilon_j)|^2-|(\epsilon_{i+1}|\epsilon_j)|^2\right)=0. \label{collision}
\end{equation}
We assume nonzero repulsion between adjacent particles, $|(\epsilon_i|\epsilon_{i+1})|\neq 0$. This leads to a condition:
\begin{equation}
\forall_{i,j} |(\epsilon_i|\epsilon_j)|^2=|(\epsilon_{i+1}|\epsilon_j)|^2\label{d2dt25}
\end{equation}
For $r>1$ all the vectors must be real, $|\epsilon_i)=|e_i)\in \mathbb{R}^r$ and in this case (\ref{d2dt25}) simplifies to
\begin{equation}
\forall_{i,j} (e_i|e_j)=\pm(e_{i+1}|e_j), 
\end{equation}
where we can choose any pair of $N-1$ nearest neighbours, thus the scalar product must be the same, up to a $\pm$ sign, for all possible pairs:
\begin{equation}
\forall_{i\neq j,k\neq l} (e_i|e_j)=\pm(e_k|e_l). \label{d2dt26}
\end{equation}
This corresponds to a system with 
\begin{equation}
L_{ij}=ig(1-\delta_{ij})e^{i\pi\cdot n_{ij}}, \hspace{5mm} n_{ij}\in\{0,1\}.\label{Lpm}
\end{equation}

In the case of $N=3$ this is equivalent to $\pm L_0$ found in (\ref{init}). It can be seen from the characteristic equation of (\ref{Lpm}):
\begin{equation}
\lambda^3-3\lambda\mp 2=(\lambda\pm 1)^2(\lambda\mp 2)=0,
\end{equation}
where the $\pm$ sign is given by $(-1)^{n_{123}}$ ($n_{ijk}=n_{ij}+n_{jk}+n_{ki}$), but regardless of the sign there is always a double eigenvalue which in case of $N=3$ corresponds to $r=3-2=1$.

Of course if all vectors $|e_i)=\pm|e)$, the condition (\ref{d2dt26}) will be satisfied for $N>3$ as well (this is again the known $r=1$ case). But what happens if there are nontrivially diffent vectors among $|e_i)$, but some of them repeat? It means that for some fixed $i,j,k$ we have $(|e_j)=\pm|e_i), (e_i|e_j)=\pm 1$ and $|e_k)\neq \pm |e_i), |(e_i|e_k)|<1$. Such a set cannot satisfy (\ref{d2dt26}), and we are left with two options: all vectors are the same up to a sign $|e_i)=\pm|e)$ (this is the case of $r=1$) or all $N$ vectors are nontrivially different, i.e. they project on different directions.

\subsubsection{The dimension of a subspace spanned by $N$ evenly distributed vectors}
We need to know what is the rank $r$ in the second case and in stead of looking at the characteristic equation we will take up a more geometrical approach. We have $N$ real vectors pointing at $N$ different, but evenly distributed directions:
\begin{equation}
\forall_{i\neq j} (e_i|e_j)=\cos(\phi_{ij})=\pm g \label{Ndirections}
\end{equation}
What is the dimension of a real space in which such a construction is possible? Of course $\mathbb{R}^N$ is enough: it contains $N$ different orthogonal vectors, which can be contracted to a narrow bundle with mutual angles smaller than $\pi/2$. This can be seen in an iterative construction for $0<g<1$:
\begin{eqnarray*}
k>l,\hspace{5mm}(e_l|e_k)&=&\sum_{i=1}^l e_l^ie_k^i+e_l^le_k^l, \hspace{5mm}
e_k^l=\frac{\sum_{i=1}^l e_l^ie_k^i}{e_l^l}\hspace{5mm},e_l^l\neq 0 \hspace{5mm}
e_k^k=\sqrt{1-\sum_{i=1}^{k-1}(e_k^i)^2}, 	\\
(e_1| &=& (1,0,0,...,0), \\
(e_2| &=& (g(-1)^{n_{12}}, \sqrt{1-g^2},0,...,0), \\
(e_3| &=& ( g(-1)^{n_{13}},g(-1)^{n_{23}}\frac{1-g(-1)^{n_{123}}}{\sqrt{1-g^2}},\sqrt{\frac{2g^3(-1)^{n_{123}}-3g^2+1}{1-g^2}},0,...0), \\
&...& \\
(e_k| &=&(g(-1)^{n_{1k}},,g(-1)^{n_{2k}}\frac{1-g(-1)^{n_{12k}}}{\sqrt{1-g^2}},...,...,...0).
\end{eqnarray*}

\begin{table}
  \begin{minipage}{\textwidth}
  	\begin{minipage}[b]{0.49\textwidth}
  		\centering
  		\includegraphics[width=7cm]{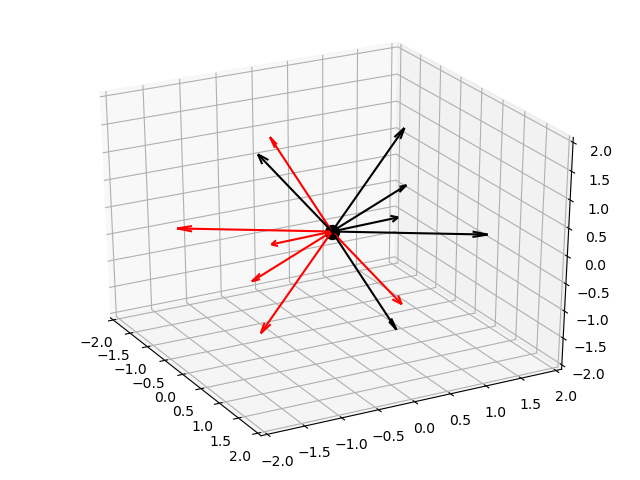}
  		\captionof{figure}{Vertices of the regular icosahedron}
  	\end{minipage}
  	\hfill
  	\begin{minipage}[b]{0.5\textwidth}
  		\centering
  		\begin{tabular}{c|c}\hline
  			N & r \\ \hline
  			3 & 1 \\
  			4 & 1,3 \\
  			5 & 1,3,4\\
  			6 & 1,3,4,5 \\
  			7 & 1,5,6 \\
  			8,9,10,11 & 1,N-3,N-2,N-1 \\

  		\end{tabular}
  		
  		\captionof{table}{Possible ranks of $N\times N$ (\ref{Lpm}) matrices calculated numerically for different combinations of $\pm$ signs. }\label{tab:ranks}
  	\end{minipage}
  \end{minipage} 
\end{table}

This is in fact the Cholesky decomposition of a positive definite matrix given by (\ref{Ndirections}) and $(e_i|e_i)=1$ on the diagonal.  As mentioned in the previous section, we can subtract $\mathbb{1}\lambda_{MIN}$ from this matrix and make it positive semidefinite. On the level of $|e_i)$ vectors this means making $e_N^N=0$ that is flattening the system of vectors to $\mathbb{R}^{N-1}$.  This means that the set of $N$ evenly distributed vectors corresponds to $r\leq N-1$. The smaller ranks are possible in higher dimensions, where for some particular configurations of $n_{ij}$ values we can make $e_l^l=0$ for $l<N$, then for all $k>l$ only $l-1$ first components of $|e_k)$ will be nonzero. This is a geometrical justification of the algebraic fact that for higher $N$ the number of classes with respect to the multiplicity of eigenvalues exceeds the two $r=1,N-1$. The possible ranks, that is the dimensions in which $N$ vectors can be packed, are presented in table \ref{tab:ranks}. In $\mathbb{R}^3$ there can be up to 6 vectors - they point to the vertices of a regular icosahedron. There are 12 of them, but once a vertex is chosen,its antipodal vertex must be excluded.
 $N\geq 8$ vectors can be distributed with $ (e_i|e_j)=\cos(\phi_{ij})=\pm g$ into $N-1,N-2$ or $N-3$ real space. An anomaly occurs for $N=7$, they fit into $\mathbb{R}^5$ and $\mathbb{R}^6$, while $\mathbb{R}^4$  accepts only up to 6 vectors.
 
\subsubsection{The second derivative of $|(\epsilon_i|\epsilon_j)|^2$ for arbitrary positions}
In the last step we look at the second time derivative (\ref{d2dt2}) in an arbitrary configuration of initial positions. Expanding (\ref{d2dt23}) with the use of (\ref{d2dt24}) and expressing the left hand side with a set of independent distances (\ref{appd2dt2}) results in a condition:
\begin{equation}
\forall_{i\neq j\neq k \neq l}(e_i|[P_l,[P_k,P_j]]|e_i) = 0.
\end{equation}
in case of real vectors satisfying (\ref{d2dt26}) this is equivalent to
\begin{equation}
\forall_{i\neq j\neq k \neq l} n_{ij}+n_{kl}=n_{ik}+n_{jl}=n_{il}+n_{kj}, \label{allijkln}
\end{equation}
which gives ${\binom{N}{4}}$ sets of equalities. In case of $N=4$ there is just one, that is: 
\begin{equation}
n_{12}+n_{34}=n_{13}+n_{24}=n_{14}+n_{23}.
\end{equation}
Keeping in mind that $n_{ij}\in\{0,1\}$ and that by addition we mean addition \emph{modulo} $2$, we can transform these equalities into
\begin{equation}
n_{12}+n_{23}+n_{31}=n_{12}+n_{24}+n_{42}=n_{13}+n_{34}+n_{41}=n_{23}+n_{34}+n_{42}.
\end{equation}
The simple relation between $n_{ij}$ and the phases $\varphi_{ij}$ defined in (\ref{ginv}), that is $\varphi_{ij}=\frac{\pi}{2}+n_{ij}\pi$, leads to the conclusion that all cyclic sums $\Phi_{ijk}$ defined by (\ref{gauge1}) must be the same, and the resulting $L$ matrix is equivalent to $L_{0,ij}=ig(1-\delta_{ij})$ found in (\ref{init}).
The generalisation of this fact onto $N>4$ is simple: the $\binom{N}{4}$ relations $\Phi_{ijk}=\Phi_{ijl}=\Phi_{ikl}=\Phi_{jkl}$ imply that $\Phi_{abc}$ is the same for all choices of $a,b,c$, and this is the definition of matrices which are equivalent to $L_{0,ij}=ig(1-\delta_{ij})$.
 This proves that the matrix found in (\ref{init}), that is the one which corresponds to $r=1$, is the only one (up to gauge equivalence) which recovers the ordinary CM model. Any higher rank $r$ allows a nontrivial unitary evolution.
 
\section{Beyond the repulsive CM model}\label{sec:beyondcm}
The vectorial formulation of the CM model contains the $(X,Y)$ Hermitian matrix dynamics as a special case. We can define a much broader set of systems in a phase space of $\{(X,Y,\mathcal{E},\mathcal{F})\}$ where X,Y are NxN matrices, while $\mathcal{E}=(|e_1),|e_2),...,|e_N))$ and $\mathcal{F}^\dagger =(|f_1),|f_2),...,|f_N))$ are rectangular matrices build of N d-dimensional column vectors \cite{gibbons84}. The symplectic form
\begin{equation}
\omega = \tr (dX\wedge dY)+i\cdot \tr (d\mathcal{F}\wedge d\mathcal{E})
\end{equation}
together with $H(X,Y,\mathcal{E},\mathcal{F})$ gives rise to the following equations of motion:
\begin{equation}
\dot{X}_{ij}=\frac{\partial H}{\partial Y_{ji}},\hspace{5mm} \dot{Y}_{ij}=-\frac{\partial H}{\partial X_{ji}},\hspace{5mm}
\dot{\mathcal{E}}_{ij}=-i\frac{\partial H}{\partial \mathcal{F}_{ji}},\hspace{5mm} \dot{\mathcal{F}}_{ij}=i\frac{\partial H}{\partial \mathcal{E}_{ji}}. \label{XYEFmotion}
\end{equation}
Without any assumptions about the $\{(X,Y,\mathcal{E},\mathcal{F})\}$ matrices we can come up with various matrix flows, which could be studied. Yet we shall assume $X,Y$ and $\mathcal{FE}$ to be Hermitian NxN matrices, since this is the case we already know how to deal with. Moreover, we restrict the possible Hamilton functions $H(X,Y,\mathcal{E},\mathcal{F})$ to those with $U(N)$ symmetry: $H(U^\dagger XU,U^\dagger YU,\mathcal{E}U,U^{\dagger}\mathcal{F})=H(X,Y,\mathcal{E},\mathcal{F})$.  The simple examples of them are the functions of $\tr (X^n),\tr (Y^n),\tr ((\mathcal{FE})^n)$. This setting allows us to perform the unitary reduction presented in section \ref{linred}. The solution $(X(t),Y(t),\mathcal{E}(t),\mathcal{F}(t))$ can be transformed (non-canonically) via the matrix which diagonalizes $X(t)$, as it was done in the case of the $(X,Y)$ system:
\begin{equation}
D(t)=U(t)X(t)U^\dagger(t),\hspace{5mm}
V(t)=U(t)Y(t)U^\dagger(t),\hspace{5mm}
E(t)=\mathcal{E}(t)U^\dagger(t),\hspace{5mm}
F(t)=U(t)\mathcal{F}(t). 
\end{equation}
This leads to the equations of motion:
\begin{equation}
\dot{D}=[A,D]+U\dot{X}U^\dagger,\hspace{5mm}
\dot{V}=[A,V]+U\dot{Y}U^\dagger, \hspace{5mm}
\dot{E}=-EA+\dot{\mathcal{E}}U^\dagger,\hspace{5mm}
\dot{F}=AF+U\dot{\mathcal{F}}, \label{EFbig}
\end{equation}
where $A=\dot{U}U^{\dagger}$.
We can express them fully with $(D,V,E,F)$ variables for a specific solution of (\ref{XYEFmotion}). 
\subsection{The CM model in the extended phase space}
The special case of Calogero-Moser Hamiltonian $H=\frac{1}{2}\tr (Y^2)$ leads, of course, to a very simple solution, trivial in $\mathcal{E},\mathcal{F}$ variables:
\begin{equation}
X(t)=X_0+t\cdot Y_0,\hspace{5mm} Y(t)=Y_0,\hspace{5mm}
\mathcal{E}(t)=\mathcal{E}_0,\hspace{5mm} \mathcal{F}(t)=\mathcal{F}_0,
\end{equation}
for which the $(D,V,E,F)$ dynamics is similar to what we already know from section \ref{linred}:
\begin{equation}
\dot{D}=[A,D]+V,\hspace{5mm}
\dot{V}=[A,V],\hspace{5mm}
\dot{L}=[A,L],\hspace{5mm} 
\dot{E}=-EA,\hspace{5mm}
\dot{F}= AF \label{DVEFFdot}
\end{equation}
This is the system (\ref{Dmdot})-(\ref{Udot}) plus the E,F degrees of freedom evolving accordingly to $A_{ij}=\frac{L_{ij}}{(D_{ii}-D_{jj})^2}$. Let us define an antihermitian matrix valued function $S$:
\begin{equation}
S = [D,V]-iFE,\hspace{5mm}\dot{S} =[A,S] 
\end{equation} 
Initial conditions $(X_0,Y_0,\mathcal{E}_0,\mathcal{F}_0)$ which satisfy $S_0 = [X_0,Y_0]-i\mathcal{E}_0\mathcal{F}_0 = -ig\cdot\mathbb{1}$ define special solutions of (\ref{DVEFFdot}) for which $S(t)=S_0$. 
What is more, these solutions correspond to the (\ref{CMef}) vectorial system:
\begin{eqnarray}
S(t)_{ii}&=&-i(f_i|e_i)(t)=-i\cdot g,\hspace{5mm}
S(t)_{ij}= L_{ij}(t)-i(f_i|e_j)(t)=0,\hspace{5mm}
V_{ij}= \frac{L_{ij}}{D_{ii}-D_{jj}} \\
\tr (V^2)&=&\sum_i V_{ii}+\sum_{i\neq j}V_{ij}V_{ji}=\sum_i p_i^2+\sum_{i\neq j}\frac{(f_i|e_j)(f_j|e_i)}{(D_{ii}-D_{jj})^2}
\end{eqnarray}
As it was shown in \ref{VecCom}, $L$ and $E,F=E^\dagger,(E^\dagger E)_{ii}$ formulations are equivalent. This fact can be expressed in the extended phase space: for $H=\frac{1}{2}\tr (Y^2)$ the flow $(X,Y,E,F)(t)$ can be restricted to orbits given by $S(0)=S_0$, on which the values of $L_{ij}$ and $(f_i|e_j)$ coincide. Other choices of $S(0)$ will result simply in a unitary evolution  $E(t)=\mathcal{E}_0 U^{\dagger}(t)$ and $F(t)=U(t)\mathcal{F}_0$ with no influence of these degrees of freedom on the $(X,Y)$ dynamics. 

\subsection{Example of interacting matrix and vectorial degrees of freedom}
Interaction between the $(X,Y)$ and vectorial degrees of freedom requires a coupling in the Hamilton function. One of the simplest examples is
\begin{equation}
H_{EF}=\frac{1}{2}\tr ((Y+\xi\mathcal{FE})^2) \label{hef}
\end{equation} 
where the free system is modified with a $\xi\mathcal{FE}$ Hermitian matrix, and $\xi^{-1}$ has the dimension of $X$. The resulting equations of motion are:
\begin{equation}
\dot{X}=Y+\xi\mathcal{FE},\hspace{5mm}\dot{Y}=0,\hspace{5mm}\dot{\mathcal{E}}=-i\xi\mathcal{E}(Y+\xi\mathcal{FE}),\hspace{5mm}\dot{\mathcal{F}}=i\xi(Y+\xi\mathcal{FE})\mathcal{F},\hspace{5mm}\dot{\Phi}=\left[i\xi Y,\Phi\right],
\end{equation}
where $\Phi=\mathcal{FE}$.
The solutions for $Y$ and $\Phi$ are straightforward, while $\mathcal{E},\mathcal{F}$ contain a nontrivial $U(d)$ modification which cancels in the $\Phi$ evolution:
\begin{equation}
Y(t)=Y_0,\hspace{5mm}\Phi(t)=e^{i\xi Y_0t}\Phi_0e^{-i\xi Y_0t},\hspace{5mm}\mathcal{E}(t)=e^{-i\xi^2\mathcal{E}_0 \mathcal{F}_0t}\mathcal{E}_0e^{-itY_0},\hspace{5mm}\mathcal{F}(t)=e^{itY_0}\mathcal{F}_0e^{-i\xi^2\mathcal{E}_0 \mathcal{F}_0t}.
\end{equation}
The equation for $X$ becomes:
\begin{equation}
\dot{X}=Y_0+\xi e^{i\xi tY_0}\Phi_0e^{-i\xi tY_0}. \label{Xef}
\end{equation}
If $[Y_0,\Phi_0]=0$, the solution is very simple: $X(t)=X_0+t(Y_0+\xi\Phi_0)$. If it is not the case, the solution reads
\begin{equation}
X(t)=X_0+tY_0+\xi\int_0^t e^{-i\xi\tau Y_0}\Phi_0 e^{-i\xi\tau Y_0}d\tau .
\end{equation}
This system turns out to have a special property:
\begin{equation}
	[X(t),Y(t)]-i\Phi(t)=	[X_0,Y_0]-i\Phi_0,
\end{equation}
i.e. $[X(t),Y(t)]-i\Phi(t)$ is a constatnt of motion, just like for the simple $Y$ dependent Hamilton function. In the next step we parametrize the flow with eigenvalues of $X(t)$ in the known way, with the use of (\ref{EFbig}), only we substitute the $E,F$ degrees of freedom with $\Omega(t)=U(t)\Phi(t)U^{\dagger}(t)$:
\begin{equation}
\dot{D}=[A,D]+V+\xi\Omega, \hspace{5mm} \dot{V}=[A,V], \hspace{5mm}\dot{\Omega}=[A+i\xi V,\Omega],\hspace{5mm}\dot{L}=[A,L]-\xi[V,\Omega], \label{DVLO}
\end{equation}
where $L=[D,V]$, and $A(t)=\dot{U}(t)U^{\dagger}(t)$ as usual. We find out that in this case $M=L-i\Omega$ satisties the same equation as $V$:
\begin{equation}
\dot{M}=[A,M], \label{MA}
\end{equation}
and therefore, just as stated in \ref{sec:intmot} for the $(D,V,L)$ flow, we have a vast set of constants of motion:
\begin{equation}
I_{k_1,k_2...,k_m}=\tr({M^{k_1}V^{k_2}\cdots M^{k_{M-1}}V^{k_M}}),
\end{equation}
where $k_1,...,k_m$ are natural numbers.
The Hamilton function (\ref{hef}) in the new $(D,V,\Omega,L)$ parametrization reads:
\begin{equation}
H_{EF}=\frac{1}{2}\tr ((V+\xi\Omega)^2)
\end{equation}
The Hamilton function in the new variables:
\begin{equation}
H_{EF}(x,p,L,\Omega)=\frac{1}{2}\sum_i p_i^2+\sum_{i<j}\frac{|L_{ij}|^2}{x^2_{ij}}-\frac{2\xi \Re(L_{ij}\Omega^*_{ij})}{|x_{ij}|}+\xi^2|\Omega_{ij}|^2
\end{equation}
is meaningful. It shows that the additional degree of freedom introduces a long distance, $1/|x|$ interaction potential between particles. The question is whether some initial conditions could recover a stationary case:
\begin{equation}
H_2(x,p)=\frac{1}{2}\sum_i p_i^2+\sum_{i<j}\frac{g_2^2}{x^2_{ij}}+\frac{g_1}{|x_{ij}|} \label{constFE}
\end{equation}
where $g_1$ can be either positive or negative. 

\subsubsection{Exact solution for $N=2$}
The simplest case, when $N=2$, can be solved analytically. Let us define the initial conditions via Pauli matrices:
\begin{equation}
X_0=x_0\sigma_z\hspace{5mm} Y_0=y_0(\bar{n}_y\cdot\bar{\sigma})\hspace{5mm}\Phi_0=\phi_0(\bar{n}_\phi\cdot\bar{\sigma}),
\end{equation}
where $\bar{n}_y,\bar{n}_\phi$ are unit vectors, and $y_0,\phi_0$ are positive constants. Note that in this definition all the matrices are traceless. The equations of motion for traces (which correspond to the center of mass motion) separate, and the equations for the traceless part reflect the relative motion. Naturally the $Y(t)$ matrix is constant, and:
\begin{eqnarray}
\Phi(t)&=&\phi_0\left[(\bar{n}_y\cdot\bar{n}_\phi)\bar{n}_y-\sin(2\xi y_0 t)(\bar{n}_y\times\bar{n}_\phi) +\cos(2\xi y_0 t)(\bar{n}_y\times\bar{n}_\phi\times\bar{n}_y) \right]\cdot\bar{\sigma}, \\
X(t)&=& X_0+t\left(1+(\bar{n}_y\cdot\bar{n}_\phi)\frac{\xi\phi_0}{y_0}\right)Y_0+ \nonumber \\
&+&\frac{\phi_0}{y_0}\sin(\xi y_0 t)\left[\sin(\xi y_0 t)(\bar{n}_y\times\bar{n}_\phi)-\cos(\xi y_0t)(\bar{n}_y\times\bar{n}_\phi\times\bar{n}_y)\right]\cdot \bar{\sigma}=\\
&=& \bar{d}(t)\cdot\bar{\sigma}.
\end{eqnarray}
The part of $\Phi_0$ which commutes with $Y_0$ contributes to the motion along the straight line, while the non-commuting part gives rise to circular motion, which means this is a spiral motion in the space of matrices. The eigenvalues, which are simply $\pm|\bar{d}(t)|$ (the distance between two particles is $r(t)=2|\bar{d}(t)|$) oscillate as well, and the oscillations are vanishing slowly, which shows that there is a long distance interaction beween particles. The Hamilton function for the relative motion in this case is:
\begin{equation}
H_2(r,p,l_x,l_y,\Omega_x,\Omega_y)=p^2+\frac{l^2_x+l^2_y}{r^2}+2\xi\frac{|l_y|\Omega_x-|l_x|\Omega_y}{r}+\xi^2(\Omega_x^2+\Omega_y^2)
\end{equation}
where $\bar{l}\cdot\bar{\sigma}=[X_0,Y_0]=2 ix_0 y_0(\hat{e}_z\times\bar{n}_y)\cdot\bar{\sigma}$ and $\bar{\Omega}\cdot\sigma  =U(t)\Phi(t)U^{\dagger}(t)$. The (\ref{constFE}) case would be recovered by such a choice of initial conditions for which:
\begin{eqnarray}
l_x^2+l_y^2 &=& const. \\
|l_y|\Omega_x-|l_x|\Omega_y &\propto& (\bar{\Omega}\times\bar{l})_z = const.
\end{eqnarray}

\begin{figure}%
	\centering
	\subfigure{{\includegraphics[width=8cm]{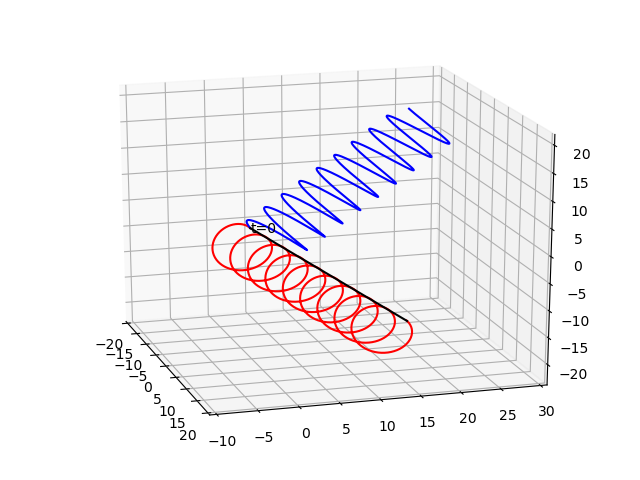} }}%
	\subfigure{{\includegraphics[width=7cm]{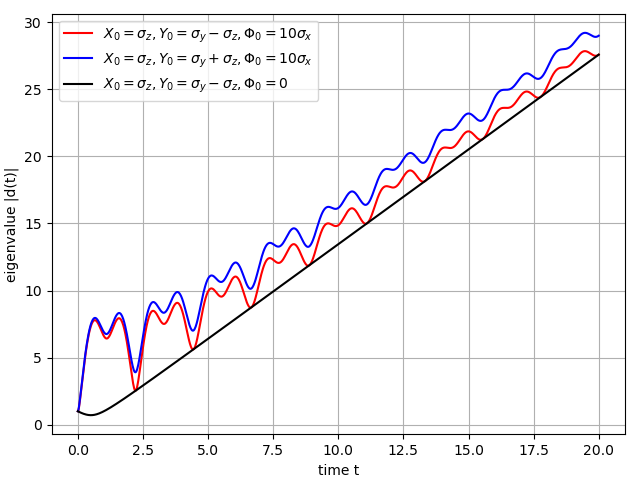} }}%
	\caption{The trajectories and lengths of $\bar{d}(t)$.The black lines represent the unextended model with $\Phi_0=0$. The legend applies to both plots.}%
	\label{fig:spirals}%
\end{figure}

Adjusting the (\ref{DVLO}) and (\ref{MA}) equations to the vectorial variables we obtain:
\begin{equation}
\dot{\bar{l}}=2i\left[(\bar{a}\times\bar{l})-\xi(\bar{v}\times\bar{\Omega})\right],\hspace{5mm}\dot{\bar{\Omega}}=2i\left[(\bar{a}+i\xi\bar{v})\times\bar{\Omega})\right],\hspace{5mm}\dot{\bar{m}}=2i(\bar{a}\times\bar{m})
\end{equation}
where $\bar{m}=i\bar{l}+\bar{\Omega}$. From a geometrical point of view we have two vectors, $\bar{\Omega}$ and $\bar{m}$ precessing (their length is constant) in $\mathbb{R}^3$ around two different directions. We require that their difference, $i\bar{l}$ maintains constant length as well. For this to be true, we need 
\begin{equation}
\frac{d}{dt}(\bar{\Omega}\cdot\bar{l})\propto (\bar{v}\times\bar{\Omega})\cdot\bar{l} = 0,
\end{equation}
as by definition $\bar{l}=2ix_0(\hat{e}_z\times\bar{v})$, and $2ix_0(v_x,v_y)=(-l_y,l_x)$, we finally obtain two possibilities:
\begin{eqnarray}
(\hat{e}_z\times\bar{v})\cdot(\hat{\Omega}\times\bar{v})&=& 0, \\
(\hat{\Omega}\times\bar{v})=0 &\vee& \frac{l^2_x+l^2_y}{4x_0^2}\Omega_z+2\xi\frac{|l_y|\Omega_x-|l_x|\Omega_y}{2x_0}v_z=0.
\end{eqnarray}
The first one, $(\hat{\Omega}\times\bar{v})=0$ means that $[Y_0,\Phi_0]=0$ which is equivalent to the non-extended $(X,Y)$ model. The second is an equation for such $\bar{l}$ and $\bar{\Omega}$ which satisfy it for any $x_0,v_z$. It implies that for $|\bar{l}|$ (the coefficient in $1/r^2$ potential)  to be constant,  $|l_y|\Omega_x-|l_x|\Omega_y$ must vanish. Yet this is the coefficient in the $1/r$ potential, which means that the only case of (\ref{constFE}) that can be obtained is when $g_1=0$, otherwise the long range part is bound to oscillate.

\subsubsection{Generalisation to higher values of $N$}
The spiral motion in the matrix space is not a unique feature of the $N=2$ case. For any value of N we cas express $X(t)$ in the diagonal basis of $Y_0$: 
\begin{equation}
(X(t))_{ij}=X_{0,ij}+t\left[\delta_{ij}(Y_{0,ii}+\xi\Phi_{0,ii})+\xi\delta(\Delta y_{ij})\Phi_{0,ij}\right]+\xi\Phi_{0,ij}\frac{f(\xi\Delta y_{ij}t)}{\Delta y_{ij}}
\end{equation} 
where $\Delta y_{ij}=Y_{0,ii}-Y_{0,jj}$ and
\begin{equation}
f(\tau)=\sin(\tau)+i(1-\cos(\tau))
\end{equation}
which represents a shifted circle in the complex plane. Therefore $X(t)-X_0$ has a linear part stemming from $Y_0$ and the part of $\Phi_0$ which commutes with $Y_0$ and a circulating part given by the entries of $[Y_0,\Phi_0]=\Delta y_{ij}\Phi_{0,ij}$. The frequency of oscillation of the $ij^{th}$ entry is $\omega_{ij}=\xi\Delta y_{ij}$. Oscillations of matrix elements will be reflected in oscillating eigenvalues, and the only $\Phi_0$ for which the oscillations vanish is the one which commutes with $Y_0$ and ads nothing but a boost to the $(X,Y)$ matrix motion. Eigenvalues of $X(t)$ are interpreted as positions of particles in an $N-body$ one-dimensional system. If the positions of the $1^{st}$ and $N^{th}$ particle oscillate in time, it means that the interaction potential continuously changes between attractive and repulsive. This implies that for a general N once we have $[Y_0,\Phi_0]\neq 0$, a stationary case (\ref{constFE}) cannot be recovered.

\section{Classical counterparts of models with spin state exchange}\label{sec:other}
The generalised Calogero-Moser models have not been quantised yet, but there are quantum extensions with spin degrees of freedom, where in the place of $L_{ij}$ variables we have matrices acting on spin states \cite{hikami93,kawakami94}. The general form of these models is the following:
\begin{equation}
\hat{\mathcal{H}}=\frac{1}{2}\sum_i p_i^2+\hbar^2\sum_{i<j}\frac{F_{ij}+E_{ij}}{\left(x_i-x_j\right)^2}
\end{equation}
where $F_{ij}=E_{ij}^2$ is diagonal in spin space, and $E_{ij}=\sum_{\alpha,\beta}g_{\alpha\beta}X_i^{\alpha\beta}X_j^{\beta\alpha}$ exchanges the spin states of $i_{th}$ and $j^{th}$ particle multiplying the state vector by a state-dependent factor. The linear term in $E_{ij}$ arrises from $\left[x_i,p_j\right]=i\hbar\delta_{ij}$ and vanishes in the classical limit .
The simplest of such models consists of pure spin state exchanges \cite{hikami93} with $g_{\alpha\beta}=1$. It can be described with operator valued matrix dynamics similar to the classical matrix model from section (\ref{linred}). A further generalization allows different weights depending on spin states \cite{kawakami94}. Although the features of quantum systems are outside the scope of this paper,we notice that the above models are not a quantum analogue of $H=\frac{1}{2}\tr (V^2)$ but of a sum of all terms:
\begin{equation}
H=\frac{1}{2}\sum_{i,j}(V^2)_{ij}
\end{equation}
which motivates us to look at the classical equations of motion for such a Hamilton function. 
\subsection{The general solution}
The classical counterpart of the discussed hamiltonians in the $(X,Y)$ phase space:
\begin{equation}
H=\frac{1}{2}\sum_{i,j} (Y^2)_{ij},
\end{equation}
results in the following equations of motion:
\begin{eqnarray}
\dot{X}_{ij}&=& Y_{ij}+\frac{1}{2}\sum_{k\neq i,j} (Y_{im}+Y_{mj}), \\
\dot{Y}_{ij}&=& 0,
\end{eqnarray}
which is again a straight line in Hermitian matrices:
\begin{eqnarray}
X(t) &=& X_0 +\frac{t}{2}[M,Y_0]_+, \\
Y(t) &=& Y_0,
\end{eqnarray}
where $M_{ij}=1$ and $[...]_+$ is the matrix anticommutator. It is much more interesting to look at the harmonic case:
\begin{equation}
H=\frac{1}{2}\sum_{i,j} (Y^2+X^2)_{ij},
\end{equation}
where the equations of motion take the following form:
\begin{eqnarray}
\dot{X}&=&\frac{1}{2}[M,Y]_+, \\
\dot{Y}&=&-\frac{1}{2}[M,X]_+.
\end{eqnarray}
Just as in the simple harmonic oscillator case, we can solve the $2^{nd}$ order differential equation for a single matrix:
\begin{equation}
\ddot{X}=-\frac{1}{4}[M,[M,X]_+]_+,\label{2ndorderX}
\end{equation}
(the same for $Y$) by means of diagonalisation. One of the possibilities is to vectorise the matrices in the equation and solve the problem using standard diagonalisation, but there is a simpler way: we notice that $M = N|e_0\rangle\langle e_0|$, where $\langle e_0|=\frac{1}{\sqrt{N}}(1,1,1,...,1)$, is proportional to a one-dimentional projection. Therefore we can express the equation (\ref{2ndorderX}) in an orthonormal basis containing $|e_0\rangle$:
\begin{eqnarray}
X &=& \left( \begin{array}{c c} x_0 & x_{i0} \\ x_{i0}^{\dagger} & x_{ij} \end{array} \right) \\
M &=& \left( \begin{array}{c c} N & 0 \\ 0 & 0 \end{array} \right) \\
\ddot{X} &=& -\frac{1}{4}[M,[M,X]_+]_+ \\
\left( \begin{array}{c c} \ddot{x}_0 & \ddot{x}_{i0} \\ \ddot{x}_{i0}^{\dagger} & \ddot{x}_{ij} \end{array} \right) &=& \left( \begin{array}{c c} -N^2x_0 & -\frac{N^2}{4}x_{i0} \\ -\frac{N^2}{4}x_{i0}^{\dagger} & 0 \end{array} \right)
\end{eqnarray}
where $x_{i0}=\langle e_0|X|e_i\rangle$ is a row vector of length $N-1$, $x_{i0}^{\dagger}$ - a column, and $x_{ij}=\langle e_i|X|e_0\rangle$ - an $(N-1)\times (N-1)$ matrix. The same procedure applies to $Y$, and with the use of the $1^{st}$ order equations: 
\begin{eqnarray}
\left( \begin{array}{c c} \dot{x}_0 & \dot{x}_{i0} \\ \dot{x}_{i0}^{\dagger} & \dot{x}_{ij} \end{array} \right) &=& \left( \begin{array}{c c} Ny_0 & \frac{N}{2}y_{i0} \\ \frac{N}{2}y_{i0}^{\dagger} & 0 \end{array} \right),\label{1storderX} \\
\left( \begin{array}{c c} \dot{y}_0 & \dot{y}_{i0} \\ \dot{y}_{i0}^{\dagger} & \dot{y}_{ij} \end{array} \right) &=& \left( \begin{array}{c c} -Nx_0 & -\frac{N}{2}x_{i0} \\ -\frac{N}{2}x_{i0}^{\dagger} & 0 \end{array} \right),
\end{eqnarray}
we find the solution:
\begin{eqnarray}
X(t) &=& \left( \begin{array}{c c} x_0\cos(Nt)+y_0\sin(Nt) & x_{i0}\cos(Nt/2)+y_{i0}\sin(Nt/2) \\ (x_{i0}\cos(Nt/2)+y_{i0}\sin(Nt/2) )^{\dagger} & x_{ij} \end{array} \right), \\
Y(t) &=& \left( \begin{array}{c c} y_0\cos(Nt)-x_0\sin(Nt) & y_{i0}\cos(Nt/2)-x_{i0}\sin(Nt/2) \\ (y_{i0}\cos(Nt/2)-x_{i0}\sin(Nt/2) )^{\dagger} & y_{ij} \end{array} \right),
\end{eqnarray}
where the $x,y_{0,i,j}$ are the elements of the initial $X_0,Y_0$ matrices in the chosen basis. We see that the $|e_0\rangle\langle e_0|$ component oscillates with the frequency $\omega = N$,the $|e_0\rangle\langle e_i|$ and $|e_i\rangle\langle e_0|$ components oscillate as well, but with a twice smaller frequency while the block which is orthogonal to $|e_0)$ stays constant. In the last step we may return to the diagonal basis of $X_0$, although then the results become less transparent - they are just linear combinations of constants and oscillating functions. 

\subsection{The $N=2$ case}
We shall illustrate the above considerations with explicit solutions for $N=2$. Let us define the evolving matrices in terms of functions, together with the initial conditions:
\begin{eqnarray}
X(t)=x_0(t)\mathbb{1}+\bar{x}(t)\cdot\bar{\sigma,}&\hspace{5mm} & X_0=x_{00}\mathbb{1}+x_{z,0}\sigma_z, \\
Y(t)=y_0(t)\mathbb{1}+\bar{y}(t)\cdot\bar{\sigma},&\hspace{5mm}& Y_0=y_{00}\mathbb{1}+\bar{y}_0\cdot\bar{\sigma}.
\end{eqnarray} 
The solution of the linear case is simple:
\begin{equation}
X(t)=X_0+t(Y_0+y_{00}\sigma_x+y_{0x}\mathbb{1}) \label{N2linearsum}
\end{equation}
As $M=\mathbb{1}+\sigma_x$, the basis which we use in the harmonic case is the eigenbasis of $\sigma_x$: $|e_0)=(1,1)/\sqrt{2},|e_1)=(1,-1)/\sqrt{2}$. The equations (\ref{2ndorderX}) and (\ref{1storderX}) in this basis read:
\begin{eqnarray}
\frac{d^2}{dt^2}\left( \begin{array}{c c} x_0(t)+x_x(t) & x_z(t)+ix_y(t) \\ x_z(t)-ix_y(t) & x_0(t)-x_x(t) \end{array} \right) &=& -\left( \begin{array}{c c} 4(x_0(t)+x_x(t)) & x_z(t)+ix_y(t) \\ x_z(t)-ix_y(t) & 0 \end{array} \right) \\
\frac{d}{dt}\left( \begin{array}{c c} x_0(t)+x_x(t) & x_z(t)+ix_y(t) \\ x_z(t)-ix_y(t) & x_0(t)-x_x(t) \end{array} \right) &=& \left( \begin{array}{c c} 2(y_0(t)+y_x(t)) & y_z(t)+iy_y(t) \\ y_z(t)-iy_y(t) & 0 \end{array} \right)
\end{eqnarray}
and the solutions are the following:
\begin{eqnarray}
x_0(t)=\frac{1}{2}\{x_{00}(1+\cos(2t))+(y_{00}+y_{0x})\sin(2t)\},&& y_0(t)=\frac{1}{2}\{(y_{00}+y_{0x})\cos(2t)-x_{00}\sin(2t)+y_{00}-y_{0x}\}, \nonumber \\
\bar{x}(t)=\left(\begin{array}{c}x_0(t)-x_{00} \\ y_{0y}\sin(t) \\ x_{0z}\cos(t)+y_{0z}\sin(t) \end{array}\right),&& \bar{y}(t)=\left(\begin{array}{c}y_0(t)-y_{00}+y_{0x} \\ y_{0y}\cos(t) \\ -x_{0z}\sin(t)+y_{0z}\cos(t) \end{array}\right).
\end{eqnarray} 
Their linear approximations coincide with (\ref{N2linearsum}). The traces, that is $\tr X(t)=2x_0(t)$ and $\tr (Y(t))=2y_0(t)$ oscillate with a period $T=\pi$ and differ only by a $\frac{\pi}{4}$ shift in time and a constant. They easily combine to a constant of motion:
\begin{equation}
(\tr X(t)-x_{00})^2+(\tr Y(t)-y_{00}+y_{0x})^2=x_{00}^2+(y_{00}-y_{0x})^2
\end{equation}
The $\bar{x}(t),\bar{y}(t)$ vectors move along periodic trajectories, which can be viewed as deformed ellipses. In particular, if the initial condition has no $\sigma_y$ component, the resulting loop will be 2-dimensional. 

The attractiveness of this model is revealed in higher dimensions, where out of all $N^2$ degrees of freedom only $N^2-(N-1)^2=2N-1$ evolve in time, while $(N-1)^2$ remain constant. Its major drawback on the other hand is such that the Hamilton functions contructed from the sum of all matrix elements instead of traces lack $U(N)$ symmetry. The unitary reduction along the lines presented in section \ref{linred} is generally impossible.

\section{Conclusions and outlook}\label{sec:conclusions}
We presented a unified approach to integrable models of the Calogero-Moser type with internal degrees of freedom. Although particular models appeared in the literature, our approach enabled us to put them all on a common ground, to show equivalences between some of them, but most importantly, to investigate the dynamics of the internal degrees of freedom, what has been rarely investigated thoroughly in previous studies. What is more, we were able to identify sets of possible values of internal degrees reachable within chosen models. The result is of significance for example in studying the level dynamics in quantum-chaotic systems, as the integral degrees of freedom play a role of coupling constants in the repulsive dynamics of energy levels. This problem will be a topic of further investigations. 

Putting all known models on a common ground of a (symplectic) reduction from a simple linear model in an extended phase space allowed constructing a new model of a similar type, reaching beyond the repulsive CM-model and investigating it from the outlined above point of view.

Last, but not least, the reduction from a linear model enables a unified and straightforward quantization of the considered systems that will be presented in a forthcoming publication by the authors.        
\appendix
\section{Details of the proofs in (\ref{rorbits})}
\subsection{Necessary conditions for (\ref{ddt})} \label{appddt}
We assume that a rational function of $N-1$ independent relative positions $x_i=x_{i1},x_{i2},..,x_{i,i-1},x_{i,i+1},..x_{ij},...,x_{iN}$:
\begin{equation}
f(x_i)=\sum_{k\neq i,j}\left(\frac{1}{x_{ik}^2}-\frac{1}{(x_{ik}-x_{ij})^2}\right)a_{ijk}
\end{equation}
is identically equal to zero. This means that for any $k\neq i,j$:
\begin{eqnarray}
\partial_{x_{ik}}f&=&\left(-\frac{2}{x_{ik}^3}+\frac{2}{(x_{ik}-x_{ij})^3}\right)a_{ijk} =2a_{ijk}\left[\frac{x_{ik}^3-(x_{ik}-x_{ij})^3}{x_{ik}^3(x_{ik}-x_{ij})^3}\right]=0 \\
\partial_{x_{ik}}f&=& 0\hspace{5mm}\iff a_{ijk}=0\vee x_{ik}=x_{ik}-x_{ij}
\end{eqnarray}
It is forbidden for $x_{ij}$ to vanish, which means $a_{ijk}=0$, and this applies to all $k\neq i,j$. In case of (\ref{ddt}) $a_{ijk}=(\epsilon_i|[P_j,P_k]|\epsilon_i)$, therefore it means that (\ref{Im0}) is the necessary condition for (\ref{ddt}) to vanish.
\subsection{Necessary conditions for (\ref{Im0})}\label{appIm0}
	The imaginary part of $(\epsilon_i|\epsilon_j)(\epsilon_j|\epsilon_k)(\epsilon_k|\epsilon_i)$ will vanish if all the vectors point in the same direction: if $|\epsilon_i) = e^{i\phi_i}|e)$ then $Im((\epsilon_i|\epsilon_j)(\epsilon_j|\epsilon_k)(\epsilon_k|\epsilon_i))=Im|e|^6=0$. If all the vectors point in one of two directions, and for the chosen triple $i,j,k$ we have for example $e^{-i\phi_i}|\epsilon_i)=e^{-i\phi_j}|\epsilon_j)=|e)$ and $e^{-i\phi_k}|\epsilon_k)=|f)$, $Im((\epsilon_i|\epsilon_j)(\epsilon_j|\epsilon_k)(\epsilon_k|\epsilon_i))=Im(|e|^2|(e|f)|^2)=0$. Three distinct directions $|\epsilon_{1,2,3}$ can be expressed as follows:
	\begin{eqnarray*}
	(\epsilon_1|&=&(1,0,0,...), \\
	(\epsilon_2|&=&(\bar{e}_2^1,\bar{e}_2^2,0,0,0...), \\
	(\epsilon_3|&=&(\bar{e}_3^1,\bar{e}_3^2,\bar{e}_k^3,0,0...),
	\end{eqnarray*} 
	and $e_{2,3}^1$ can be made real by a correct choice of gauge. The basis independent value of (\ref{Im0}) reads:
	\begin{equation}
	Im(e_2^1(e_2^1e_3^1+\bar{e}_2^2e_2^2)e_3^1)=0
	\end{equation} 
	which is equivalent to $Im(\bar{e}_2^2e_3^2)=0$, meaning that $e_2^2$ and $e_3^2$ have a common phase factor. Applying the same procedure to all tripples of vectors, but expressing everything in the initial basis, we find out that
		\begin{eqnarray*}
			(\epsilon_1|&=&(1,0,0,...), \\
			(\epsilon_2|&=&(e_2^1,e^{i\phi_2}e_2^2,0,0,0...), \\
			(\epsilon_3|&=&(e_3^1,e^{i\phi_2}e_3^2,e^{i\phi_3}e_3^3,0,0...)\\
			(\epsilon_4|&=&(e_4^1,e^{i\phi_2}e_4^2,e^{i\phi_3}e_4^3,e^{i\phi_4}e_4^4,0...),\\
			&..&\\
			(\epsilon_N|&=&(e_N^1,e^{i\phi_2}e_N^2,e^{i\phi_3}e_N^3,...,e^{i\phi_N}e_N^N),
		\end{eqnarray*} 
where all the $e_k^l\in\mathbb{R}$, and the resulting $L_{ij}=i(\epsilon_i|\epsilon_j)$ matrix is equal to the one given by purely real vectors without the phase factors.
\subsection{Necessary conditions for (\ref{d2dt23})}\label{appd2dt2}
Let us find the necessary condition for:
\begin{equation}
\sum_{k\neq i,j}\left(\frac{1}{x_{ik}^2}-\frac{1}{x_{jk}^2}\right)\left(\sum_{l\neq i}\frac{1}{x_{il}^2}(\epsilon_i|[P_l,[P_k,P_j]]|\epsilon_i)+\sum_{l\neq j}\frac{1}{x_{jl}^2}(\epsilon_i|[P_k,[P_j,P_l]]|\epsilon_i)+\sum_{l\neq k}\frac{1}{x_{kl}^2}(\epsilon_i|[P_j,[P_l,P_k]]|\epsilon_i)\right)=0.
\end{equation}
provided that the vectors are real $|\epsilon_i)=|e_i)\in R^r$, and as a concequence $(e_a|e_b)=(e_b|e_a)$. First of all we shall introduce convenient shorthands: 
\begin{eqnarray}
[abc]=-[acb]=(e_i|[P_a,[P_b,P_c]]|e_i)&=& 2(e_i|e_a)(e_b|e_c)((e_a|e_b)(e_c|e_i)-(e_a|e_c)(e_b|e_i)), \label{Pabc} \\
\left[abc\right]+[bca]+[cab] &=& 0,  \\
(ab)=(ba)&=&(e_a|e_b).
\end{eqnarray}
The sums over $l$ exclude just one index each, for example $i$, but allow $j,k$, and therefore we have:
\begin{eqnarray}
\sum_{k\neq i,j}\left(\frac{1}{x_{ik}^2}-\frac{1}{x_{jk}^2}\right)\left(\frac{[jkj]+[kji]}{x_{ij}^2}+\frac{[kkj]+[jik]}{x_{ik}^2}+\frac{[kjk]+[jjk]}{x_{jk}^2}\right)+& & \label{kij}\\
+\sum_{k\neq i,j}\left(\frac{1}{x_{ik}^2}-\frac{1}{x_{jk}^2}\right)\left(\sum_{l\neq i,j,k}\frac{[lkj]}{x_{il}^2}+\frac{[kjl]}{x_{jl}^2}+\frac{[jlk]}{x_{kl}^2}\right)= & 0 &.
\end{eqnarray} 
We use the righthand side of (\ref{Pabc}) to express the $[abc]$ commutators in the (\ref{kij}) part of the equation:
\begin{eqnarray}
2\sum_{k\neq i,j}\left(\frac{1}{x_{ik}^2}-\frac{1}{x_{jk}^2}\right)\left(\frac{(ij)^2((jk)^2-(ik)^2)}{x_{ij}^2}+\frac{(ik)^2((ij)^2-(jk)^2)}{x_{ik}^2}+\frac{(jk)^2((ik)^2-(ij)^2}{x_{jk}^2}\right)&+& \label{kij2}\\
+\sum_{k\neq i,j}\left(\frac{1}{x_{ik}^2}-\frac{1}{x_{jk}^2}\right)\left(\sum_{l\neq i,j,k}\frac{[lkj]}{x_{il}^2}+\frac{[kjl]}{x_{jl}^2}+\frac{[jlk]}{x_{kl}^2}\right)= & 0 &. \label{klij}
\end{eqnarray} 
The terms which are proportional to $x_{ik}^{-4},k=i+1$ are dominant in the colliding pair approximation. Here it is visible how the (\ref{collision}) condition arises from the equation for general $\bar{x}=(x_{i,1},x_{i,2},...,x_{i,N})$. The (\ref{kij2}) part will vanish for $(ab)=(e_a|e_b)=g(-1)^{n_{ab}}$. Now we have to check the conditions for the (\ref{klij}) part in this case (using the Jacobi identity $[jkl]=-[lkj]-[kjl]$):
\begin{eqnarray}
(e_a|e_b)&=& g(-1)^{n_{ab}} \Phi_{iabc}=(-1)^{n_{ia}+n_{ab}+n_{bc}+n_{ci}}, \\ \left[abc\right]&=& 2g^4(-1)^{n_{ia}+n_{bc}}((-1)^{n_{ab}+n_{ic}}-(-1)^{n_{ac}+n_{ib}})=2g^4(\Phi_{iabc}-\Phi_{iacb}), \\
0&=&2\sum_{k\neq l \neq i,j}\left(\frac{1}{x_{ik}^2}-\frac{1}{x_{jk}^2}\right)\left(\left(\frac{1}{x_{il}^2}-\frac{1}{x_{kl}^2}\right)(\Phi_{ilkj}-\Phi_{iljk})+\left(\frac{1}{x_{jl}^2}-\frac{1}{x_{kl}^2}\right)(\Phi_{ikjl}-\Phi_{iklj})\right).
\end{eqnarray}
In the next step we modifying the sum so that we have all $k,l$ dependent terms grouped together:
\begin{eqnarray}
0&=&\sum_{k<l \neq i,j}\left(\frac{1}{x_{ik}^2}-\frac{1}{x_{jk}^2}\right)\left(\left(\frac{1}{x_{il}^2}-\frac{1}{x_{kl}^2}\right)(\Phi_{ilkj}-\Phi_{iljk})+\left(\frac{1}{x_{jl}^2}-\frac{1}{x_{kl}^2}\right)(\Phi_{ikjl}-\Phi_{iklj})\right)+\\
&+&\left(\frac{1}{x_{il}^2}-\frac{1}{x_{jl}^2}\right)\left(\left(\frac{1}{x_{ik}^2}-\frac{1}{x_{kl}^2}\right)(\Phi_{iklj}-\Phi_{ikjl})+\left(\frac{1}{x_{jk}^2}-\frac{1}{x_{kl}^2}\right)(\Phi_{iljk}-\Phi_{ilkj})\right)=\\
&=&\sum_{k<l \neq i,j}A_{ijkl}\left[\left(\frac{1}{x_{ik}^2}-\frac{1}{x_{jk}^2}\right)\left(\frac{1}{x_{il}^2}-\frac{1}{x_{kl}^2}\right)-\left(\frac{1}{x_{il}^2}-\frac{1}{x_{jl}^2}\right)\left(\frac{1}{x_{jk}^2}-\frac{1}{x_{kl}^2}\right)\right]+\\
&+& B_{ijkl}\left[\left(\frac{1}{x_{ik}^2}-\frac{1}{x_{jk}^2}\right)\left(\frac{1}{x_{jl}^2}-\frac{1}{x_{kl}^2}\right)-\left(\frac{1}{x_{il}^2}-\frac{1}{x_{jl}^2}\right)\left(\frac{1}{x_{ik}^2}-\frac{1}{x_{kl}^2}\right)\right],
\end{eqnarray}
where $A_{ijkl}=\Phi_{ilkj}-\Phi_{iljk}$ and $B_{ijkl}=\Phi_{ikjl}-\Phi_{iklj}$. If we parmetrise the expression with $N-1$ independent distances $x_{i1},x_{i2},...x_{i,i-1},x_{i,i+1},...,x_{ij},..x_{iN}$, we can write:
\begin{equation}
0 = \sum_{k<l}A_{ijkl}F(x_{ij},x_{ik},x_{il})+B_{ijkl}G(x_{ij},x_{ik},x_{il}), \label{sum105}
\end{equation}
where 
\begin{eqnarray}
F(x,y,z)&=& f(y,y-x)f(z,y-z)-f(z,z-x)f(y-x,z-y),\\
G(x,y,z)&=& f(y,y-x)f(z-x,y-z)-f(z,z-x)f(y,y-z), \\
f(a,b)&=&\frac{1}{a^2}-\frac{1}{b^2}.
\end{eqnarray}
In case of $i=1,j=2$ and $N=4$ this sum has only one term corresponding to $k=3,l=4$ and after collecting the terms in the numerator we obtain the condition
\begin{eqnarray}
0 &=&-6(A+B)x_{13}^2 x_{14}^2 +2x_{12} ((5 A + B) x_{13}^2 x_{14} + (A + 5 B) x_{13} x_{14}^2)+\nonumber \\
& &-4x_{12}^2 (A x_{13}^2 + (A+B) x_{13} x_{14}+B x_{14}^2) +  2x_{12}^3 (A x_{13} + B x_{14}), \label{1234cond}
\end{eqnarray}
where $A=A_{1234},B=B_{1234}$ and as both $A,B\in\{0,\pm 2\}$, we see that only $A=B=0$ can satisfy this condition for all $x_{12}<x_{13}<x_{14}$. This means that $(e_1|[P_4,[P_3,P_2]]|e_1)=(e_1|[P_3,[P_4,P_2]]|e_1)=0$, and $n_{13}+n_{24}=(n_{14}+n_{23}) \emph{mod} 2$. By choosing another $(i,j)$ pair we obtain the equality of $n_{12}+n_{34}=n_{13}+n_{24}=(n_{14}+n_{23}) \emph{mod} 2$. To extend this result to $N>4$  without dealing with multiple terms,we notice that for fixed $m\neq n\neq i$ only one term in  (\ref{sum105}) depends on both $x_{im}$ and $x_{in}$. Since (\ref{sum105}) is required to be identically zero, its derivatives must vanish as well. Therefore for a fixed pair $m,n$:
\begin{eqnarray}
0 &=& \partial^2_{x_{im}x_{in}}\left(\sum_{k<l}A_{ijkl}F(x_{ij},x_{ik},x_{il})+B_{ijkl}G(x_{ij},x_{ik},x_{il})\right)=\\
&=&A_{ijmn}\partial^2_{x_{im}x_{in}}F(x_{ij},x_{im},x_{in})+B_{ijmn}\partial^2_{x_{im}x_{in}}G(x_{ij},x_{im},x_{in}),
\end{eqnarray}
which leads to a condition 
\begin{eqnarray*}
0&=& 
x^5 (4 A y^3 + (-9 A + 3 B) y^2 z + (3 A - 9 B) y z^2 + 4 B z^3) + \\
& & +x^4 (-12 A y^4 + (15 A - 9 B) y^3 z + 18(A+B) y^2 z^2 + (-9 A + 15 B) y z^3 - 12 B z^4) + \\
& & +x^3 (12 A y^5 +9(A+B) y^4 z - 60 A y^3 z^2 - 
60 B y^2 z^3 + 9(A+B) y z^4 + 12 B z^5) + \\
& &+x^2 ((-39 A - 3 B) y^5 z + (54 A - 18 B) y^4 z^2 + 44(A + B) y^3 z^3 + (-18 A + 54 B) y^2 z^4 + (-3 A - 39 B) y z^5) +\\
& & +x ((45 A + 9 B) y^5 z^2 + (-93 A + 15 B) y^4 z^3 + (15 A - 
93 B) y^3 z^4 + (9 A + 45 B) y^2 z^5)+\\
& &+(A+B)(-18y^5 z^3 + 42y^4 z^4 -18y^3 z^5) 
\end{eqnarray*}
where $A=A_{ijmn},B=B_{ijmn},x=x_{ij},y=x_{im},z=x_{in}$. It is satisfied for all possible distances only if $A_{ijmn}=B_{ijmn}=0$, that is if (\ref{allijkln}) is satisfied.

\section*{Acknowledgments}

The authors acknowledge a financial support of the the Polish National Science Centre \\ grant 2017/27/B/ST2/02959.


\begin{thebibliography}{10}
	
	\bibitem{calogero69}
	F.~Calogero.
	\newblock {Solution of a Three-Body Problem in One Dimension}.
	\newblock {\em J. Math. Phys.}, 10(12):2191--2196, 1969.
	
	\bibitem{calogero69a}
	F.~Calogero.
	\newblock {Ground State of a One-Dimensional N-Body System}.
	\newblock {\em J. Math. Phys.}, 10(12):2197--2200, 1969.
	
	\bibitem{calogero71}
	F.~Calogero.
	\newblock {Solution of the One-Dimensional N-Body Problems with Quadratic
		and/or Inversely Quadratic Pair Potentials}.
	\newblock {\em J. Math. Phys.}, 12(3):419--436, 1971.
	
	\bibitem{olshanetsky83}
	M.~A. Olshanetsky and A.~M. Perelomov.
	\newblock {Quantum integrable systems related to Lie algebras}.
	\newblock {\em Phys. Rep.}, 94(6):313--404, 1983.
	
	\bibitem{olshanetsky81}
	M.~A. Olshanetsky and A.~M. Perelomov.
	\newblock {Classical integrable finite-dimensional systems related to Lie
		algebras}.
	\newblock {\em Phys. Rep.}, 71(5):313--400, 1981.
	
	\bibitem{sutherland71}
	B.~Sutherland.
	\newblock {Exact Results for a Quantum Many-Body Problem in One Dimension}.
	\newblock {\em Phys. Rev. A}, 4:2019--2021, 1971.
	
	\bibitem{sutherland72}
	B.~Sutherland.
	\newblock {Exact Results for a Quantum Many-Body Problem in One Dimension. II}.
	\newblock {\em Phys. Rev. A}, 5:1372--1376, 1972.
	
	\bibitem{calogero75}
	F.~Calogero, O.~Ragnisco, and C.~Marchioro.
	\newblock Exact solution of the classical and quantal one-dimensional many-body
	problems with the two-body potential $v_a(x)=g^2a^2/\mathrm{sh}^2(ax)$.
	\newblock {\em Lettere al Nuovo Cimento}, 13:383--387, 1975.
	
	\bibitem{gibbons84}
	J.~Gibbons and T.~Hermsen.
	\newblock {A generalisation of the {C}alogero-{M}oser system}.
	\newblock {\em {Physica D: Nonlinear Phenomena}}, 11(3):337 -- 348, 1984.
	
	\bibitem{ha92}
	Z.~N.~C. Ha and F.~D.~M. Haldane.
	\newblock Models with inverse-square exchange.
	\newblock {\em Phys. Rev. B}, 46:9359--9368, Oct 1992.
	
	\bibitem{hikami93}
	K.~Hikami and M.~Wadati.
	\newblock Infinite symmetry of the spin systems with inverse square
	interactions.
	\newblock {\em J. Phys. Soc. Japan}, 62(12):4203--4217, 1993.
	
	\bibitem{minahan93}
	J.~A. Minahan and A.~P. Polychronakos.
	\newblock {Integrable Systems for Particles with Internal Degrees of Freedom}.
	\newblock {\em Phys. Lett. B}, 302(2):265--270, 1993.
	
	\bibitem{krichever95}
	I.~Krichever, O.~Babelon, E.~Billey, and M.~Talon.
	\newblock {Spin generalization of the Calogero-Moser system and the matrix KP
		equation}.
	\newblock {\em {Translations of the American Mathematical Society-Series 2}},
	170:83--120, 1995.
	
	\bibitem{kawakami94}
	N.~Kawakami and Y.~Kuramoto.
	\newblock Hierarchical models with inverse-square interaction in harmonic
	confinement.
	\newblock {\em Phys. Rev. B}, 50(7):4664--4670, 1994.
	
	\bibitem{haake19}
	F.~Haake, S.~Gnutzmann, and M.~Ku\'s.
	\newblock {\em {Quantum Signatures of Chaos}}.
	\newblock Springer, International Publishing, 2019.
	
	\bibitem{moser75}
	J.~Moser.
	\newblock {Three Integrable Hamiltonian Systems Connected with Isospectral
		Deformations}.
	\newblock {\em Adv. Math.}, 16(2):197--220, 1975.
	
	\bibitem{kus97}
	M.~Ku\'s, F.Haake, D.~Zaitsev, and A.~Huckleberry.
	\newblock {Level dynamics for conservative and dissipative quantum systems}.
	\newblock {\em J. Phys. A: Math. Gen.}, 30(24):8635--8651, 1997.
	
	\bibitem{huckleberry08}
	A.~Huckleberry, M.~Ku\'s, and P.~Sch{\"u}tzdeller.
	\newblock {Level dynamics and the ten-fold way}.
	\newblock {\em J. Geom. Phys.}, 58:1231--1240, 2008.
	
	\bibitem{abraham78}
	R.~Abraham and J.~Marsden.
	\newblock {\em {Foundations of Mechanics}}, volume~36.
	\newblock Benjamin/Cummings Publishing Company Reading, Massachusetts, 1978.
	
	\bibitem{huckleberry01}
	A.~Huckleberry, D.~Zaitsev, M.~Ku\'s, and F.~Haake.
	\newblock {A symplectic context for level dynamics}.
	\newblock {\em J. Geom. Phys.}, 37:156--168, 2001.
	
	\bibitem{kirillov04}
	A.~A. Kirillov.
	\newblock {\em {Lectures on the Orbit Method}}.
	\newblock American Mathematical Soc., 2004.
	
	\bibitem{mnich93}
	K.~Mnich.
	\newblock {A complete set of constants of motion for the generalized
		Sutherland-Moser system}.
	\newblock {\em Phys. Lett. A}, 176(3-4):189--192, 1993.
	
	\bibitem{jurdjevich97}
	V.~Jurdjevich.
	\newblock {\em {Geometric Control Theory}}.
	\newblock Cambridge University Press, 1997.
	
\end{thebibliography}

\end{document}